# Revealing the Staging Structural Evolution and Li (De)Intercalation Kinetics in Graphite Anodes via Machine Learning Potential

Liqi Wang[a, b], Xuhe Gong[a, c], Zicun Li[a], Ruijuan Xiao[a, b, *] and Hong Li[a, b, *]

a Institute of Physics, Chinese Academy of Sciences, Beijing, 100190, China.

b School of Physical Sciences, University of Chinese Academy of Sciences, Beijing 100049, China

c School of Materials Science and Engineering, Key Laboratory of Aerospace Materials and Performance (Ministry of Education), Beihang University, Beijing 100191, China.

*Corresponding author. Email: rjxiao@iphy.ac.cn, hli@iphy.ac.cn

**ABSTRACT**

Revealing the dynamic structural evolution and lithium transport properties during the charge/discharge processes is crucial for optimizing graphite anodes in lithium-ion batteries, enabling high stability and fast-charging performance. However, the dynamic coupling mechanisms among carbon layer kinetics, lithium (de)intercalation/diffusion, and defects regulation remain insufficiently understood. In this study, we developed a universal automated workflow based on machine learning potentials to simulate the dynamic lithium (de)intercalation process. With this approach, the staging structural evolution of lithium-graphite intercalation compounds and their lithium transport behavior were resolved through molecular dynamics simulations. By introducing stacking faults into the graphite structure, we successfully simulated stage transitions driven by carbon layer sliding and reorganization, accompanied by stress release and structural stabilization. The dynamics of carbon layers regulate the lithium (de)intercalation positional selectivity, producing intermediate states with varying lithium concentrations and distributions during cycling. This facilitates the formation and transformation of stage structures while mitigating residual stress accumulation. A fundamental kinetic asymmetry arises between lithium intercalation and deintercalation, driven by the continuous and heterogeneous lithium transport and carbon layer sliding during charge/discharge processes. The carbon defects regulate lithium transport behavior, in which the atomic-scale defects confine intralayer lithium transport and carbon sliding while enabling interlayer transport via dynamic lithium trapping/release mechanisms, and atomic-layer-scale defects concurrently boosting lithium diffusivity and ameliorating structural stress distributions.

Accordingly, for the future graphite anode design, it is critical to construct structural units with controllable carbon layer sliding and reorganization behaviors, as well as tunable defects to enhance lithium-ion transport, mitigate structural evolution asymmetry, and maintain the mechanical stability.

**INTRODUCTION**

Lithium-ion batteries (LIBs) are pivotal in contemporary energy storage systems, with current research efforts primarily focusing on enhancing the cycling stability and fast-charging capability.[1-6] The graphite anode, due to its low operating potential (approximately 0.1 V vs. Li$^+$/Li) and good structural stability, has become the most widely used anode material in LIBs.[7-14] During the charge/discharge cycles, the insertion and extraction of Li between C layers induce complex phase transition behaviors, forming various stage structures from Stage-1 (LiC$_6$) to Stage-4 (LiC$_{24}$), accompanied by the generation of defects.[15-21] Unresolved issues—including the formation mechanisms of stage structures, the kinetics of stage transitions, and the origins of hysteresis in charge/discharge profiles—continue to constrain the graphite anode stability and its fast-charging application.[22-24] Additionally, the presence of defect structures can affect Li transport behavior, leading to irreversible capacity loss and affecting charge-discharge efficiency.[25-26] Therefore, a multiscale understanding is needed to elucidate the impact of Li-GICs microstructure evolution on Li transport, specifically including the formation mechanisms and transition processes of stage structures, the evolution of lithium transport behavior and kinetics of carbon layers during charge/discharge cycles, and the regulatory effects of defect structures on lithium transport behavior.

To understand the formation of stage structures, two classical models have emerged: the Rüdorff-Hofmann model and the Daumas-Hérold model.[27-28] The former suggests that Li alternately inserts into the gaps between C layers; for an n-stage structure, there are n-1 C layers between two adjacent Li layers. Due to kinetic limitations, transitions between these structures are difficult, hindering the fast charging capability of graphite.[29] The latter model proposes that during Li intercalation, the C layers undergo deformation, which facilitates the transformation between stage structures.[30] Both models reflect the long-range order of stage structures, but there are differences in their microscopic descriptions and the intermediate states during stage structure transitions. Later, to reveal the short-range reality, Weng et al. observed the microstructure of Li-GICs using cryo-transmission electron microscopy (cryo-TEM), which is sensitive to microstructural changes, and proposed the localized-domains model.[25] This model posits that during Li insertion, the C layers gradually become disordered, forming numerous dislocations and microdomains that contain different stage structures. The deformation of C layers is attributed to local stress caused by the uneven distribution of Li. This model retains the long-range order while unveiling the microscopic heterogeneity of Li-GICs, providing new insights into the formation of stage structures. On the theoretical simulation front, due to the complexity of the phase transition processes in Li-GICs which poses challenges to traditional DFT and AIMD approaches, Baber et al., Yang et al., and Panosetti et al. independently adopted machine learning potential methods to simulate voltage profiles, energy and specific capacity variations during (de)lithiation, as well as structural information, thereby providing several valuable insights.[31-33] The experimental and theoretical methods employed in these research efforts capture the structural features of Li-GICs, providing essential support for understanding the formation, structural evolution, and Li transport characteristics of stage structures. However, due to the limitations of current experimental techniques in spatiotemporal resolution, as well as the constraints of theoretical studies in covering structural evolution scenarios, simulation time and spatial scales, and the ability to capture non-equilibrium states, the atomistic mechanisms governing stage structure formation, their controlling factors, the influence of defects on lithium-ion transport behavior remain insufficiently understood. To gain a more intuitive understanding of the dynamic structural evolution of Li-GICs and the mechanisms of lithium (de)intercalation, it is necessary to achieve experimental resolutions capable of capturing sub-second transformation processes or to conduct theoretical simulations that span the entire transformation timescale and incorporate non-equilibrium states.

In recent years, the development of the Deep Potential (DP) model has provided new methods for addressing these challenges[34-36]. This model uses neural networks to fit high-dimensional potential surfaces, significantly enhancing the speed of molecular dynamics (MD) simulations while maintaining quantum mechanical accuracy. This advantage makes long-term dynamic simulations of large-scale nanostructures (approximately 10,000 atoms, ~10 ns) feasible, providing a powerful tool for analyzing the structural evolution of lithium graphite intercalation compounds (Li-GICs)[37-38]. In this study, we constructed a DP model for Li-GICs and combined it with large-scale, long-duration molecular dynamics simulations (DPMD) to overcome the space-time limitations of traditional methods. Our approach demonstrates enhanced accuracy in energy and atomic force predictions on test sets while expanding research dimensions including full-cycle electrochemical simulations capturing complete charge/discharge processes, stage transformational dynamics during phase transitions, and defect-mediated regulation of Li transport and C layer kinetics, among other multifaceted research capabilities. Through using Large-scale Atomic/Molecular Massively Parallel Simulator (LAMMPS)[39], we employed our model to conduct DPMD simulations of Li-GICs across 4 key research dimensions. First, building upon experimental observations of stacking fault defects in C layers, we established computational models and verified through DPMD simulations that sliding and reorganization of C layers can drive stage structure transformations. Second, by conducting comparative full-cycle charge/discharge simulations between pristine C structures and those with single-atom vacancies, we systematically tracked dynamic evolution in Li transport capabilities and C layer kinetic behavior during Li intercalation/deintercalation processes while observing inherent process asymmetry. Subsequently, leveraging structural dynamic evolution captured in these simulations, we investigated positional selectivity in Li (de)intercalation governed by C layer kinetics and further analyzed how resultant Li concentration gradients and spatial distribution patterns modulate diffusion kinetics and C layer response. Finally, through probing atom-scale defect and atom-layer-scale defects, we systematically evaluated differential impacts of multiscale defects on Li transport efficiency and C layer stability. These studies deepen the understanding of the multi-scale dynamic properties of Li-GICs, providing theoretical guidance for designing stable and fast-charging graphite anodes.

**RESULT AND DISSCUSSION**

To gain a deeper understanding of the complex dynamic behavior of lithium graphite intercalation compounds (Li-GICs), we constructed a dataset encompassing different pressures, Li concentrations, various staging structures, and defect structures (details on dataset preparation, model training parameter settings, model testing, and further validation can be found in the Supporting Information under the section titled *DP Model Training, Testing and Further Validation*). After multiple rounds of iterative training and rigorous validation, we ultimately developed a machine learning potential model for Li-GICs (hereafter referred to as the DP model). As shown in Fig. 1a, with the DP model in hand, we simulated the stage structure transformation process based on C layer sliding and reorganization, as well as the complete charge/discharge processes. Furthermore, we investigated the selectivity of Li insertion/extraction sites, the effects of Li concentration and distribution on Li transport and C layer dynamics, and the regulatory roles of defect structures in Li transport and C layer dynamics.

To evaluate the basic predictive capability (Tab. S1) of the DP model for different Li-GICs structures, we compared the DP prediction results with DFT calculated results based on the test set, assessing the model's accuracy in predicting energy, atomic forces, and the system's equation of state. As shown in Figs. 1b and 1c, the red data points representing the model predictions are tightly clustered around the blue diagonal line which indicates perfect agreement with the DFT results. Specifically, the mean absolute error (MAE) for energy predictions is 2.787 meV/atom, and the root mean square error (RMSE) is 5.276 meV/atom. The MAE and RMSE values for atomic force predictions are 125 meV/Å and 238 meV/Å, respectively. As shown in Figs. S1a and S1b, for external pressures ranging from -50 kbar to 50 kbar, the maximum prediction error of both AA-stacked and AB-stacked configurations is less than 7 meV/atom.

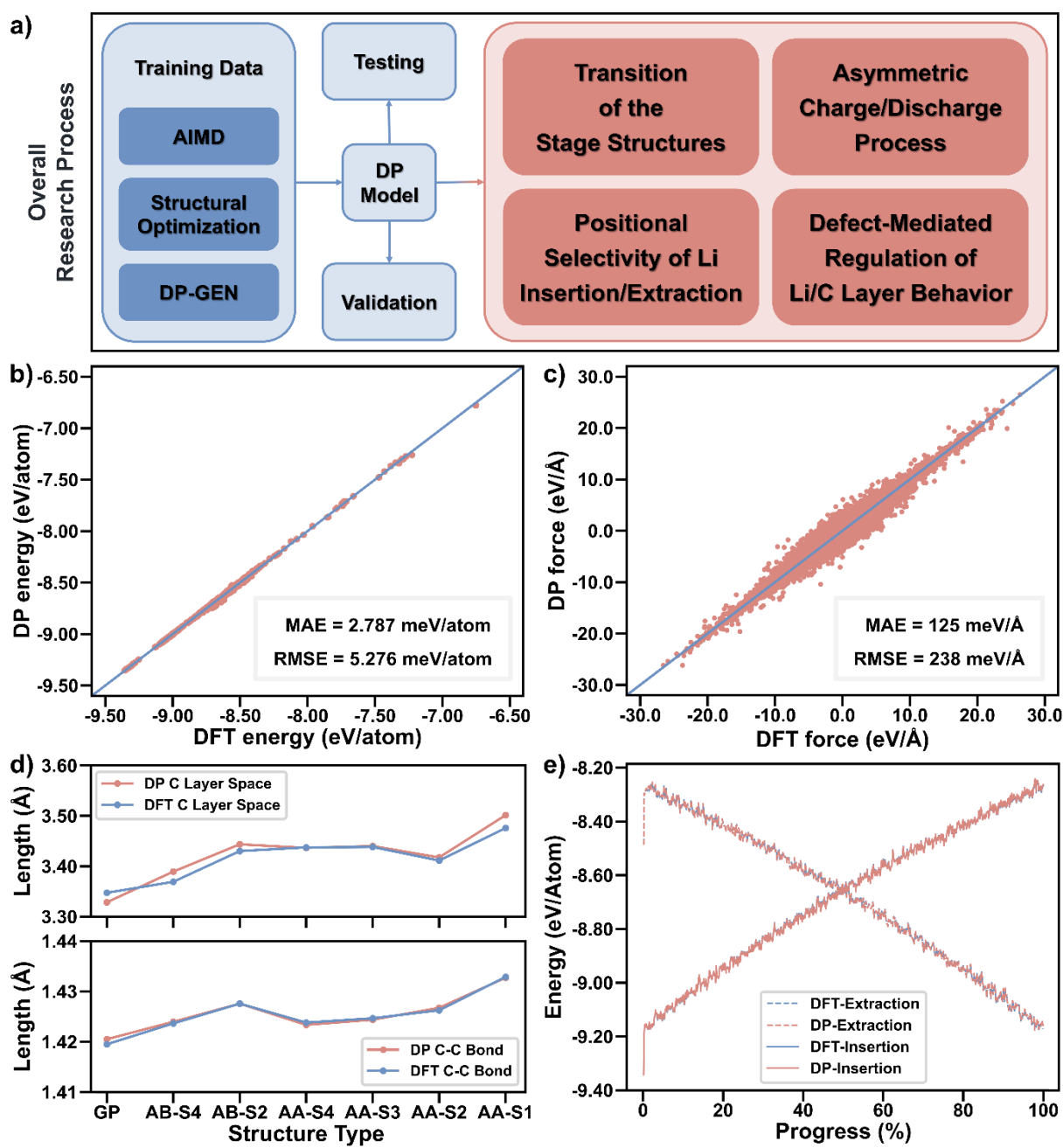

Figure 1: Overall Research Process and DP Model Testing and Validation for Li-GICs in This Work (a) The research process of this work. (b) Evaluation of the energy prediction accuracy of the DP model based on the test set. (c) Evaluation of the atomic force prediction accuracy of the DP model based on the test set. (d) Comparison of the average C layer spacings (upper panel) and average C-C bond lengths (lower panel) predicted by the DP model with DFT results for various structures (from left to right: graphite pristine, AB-stacked Stage-4, AB-stacked Stage-2, AA-stacked Stage-4, AA-stacked Stage-3, AA-stacked Stage-2, and AA-stacked Stage-1 structures). (e) Validation of the consistency in predictive accuracy of DPMD simulations performed with the DP model at different lithium concentrations, based on single-ion-step DFT

calculations. The data composition for model testing and further validation is provided in Tab. S1.

To further validate (Tab. S1) the capability of the DP model in describing structural evolution and dynamic processes during charge and discharge, we compared the structural feature dimensions predicted by the DP model with the results from DFT calculations. Additionally, we compared the energy variation given by DPMD simulations for the Li insertion/extraction process with DFT results. As shown in Fig. 1d, the predicted values and trends for the average C layer spacings and average C-C bond lengths exhibit good consistency with the DFT results, with a maximum deviation of only 0.047 Å. This indicates that even as the Li concentration in the structure continuously changes, the predictive accuracy of the DP model remains stable, further demonstrating the good adaptability of the DP model to simulation scenarios with varying Li concentrations.

Benefitting from the above evaluation and validation results, we believe that the DP model has the capability to accurately describe the fundamental properties of different Li-GICs structures and simulate their dynamic evolution. Even when external pressures and the Li concentration within the structure change, the model still demonstrates reliable predictive results. In the following sections, we present new insights into Li-GICs enabled by DPMD simulations, revealing previously inaccessible aspects such as the formation and transformation of staging structures in graphite, as well as the structural evolution and lithium transport behavior throughout the full charge/discharge process.

1. **Transition of Stage Structures**

 After validating the DP model, we first investigated the stage structure transition processes of Li-GICs. Among various models[25, 27, 28], the key issues involve the deformation and local reorganization of C layers, as well as their role in stage formation and transformation. Therefore, we constructed multi-stage structural models (from stage-1 to stage-4) with different numbers of stacking faults and monitored their dynamic evolution through DPMD simulations (1200 K, 6312-7368 atoms, 100 ps).

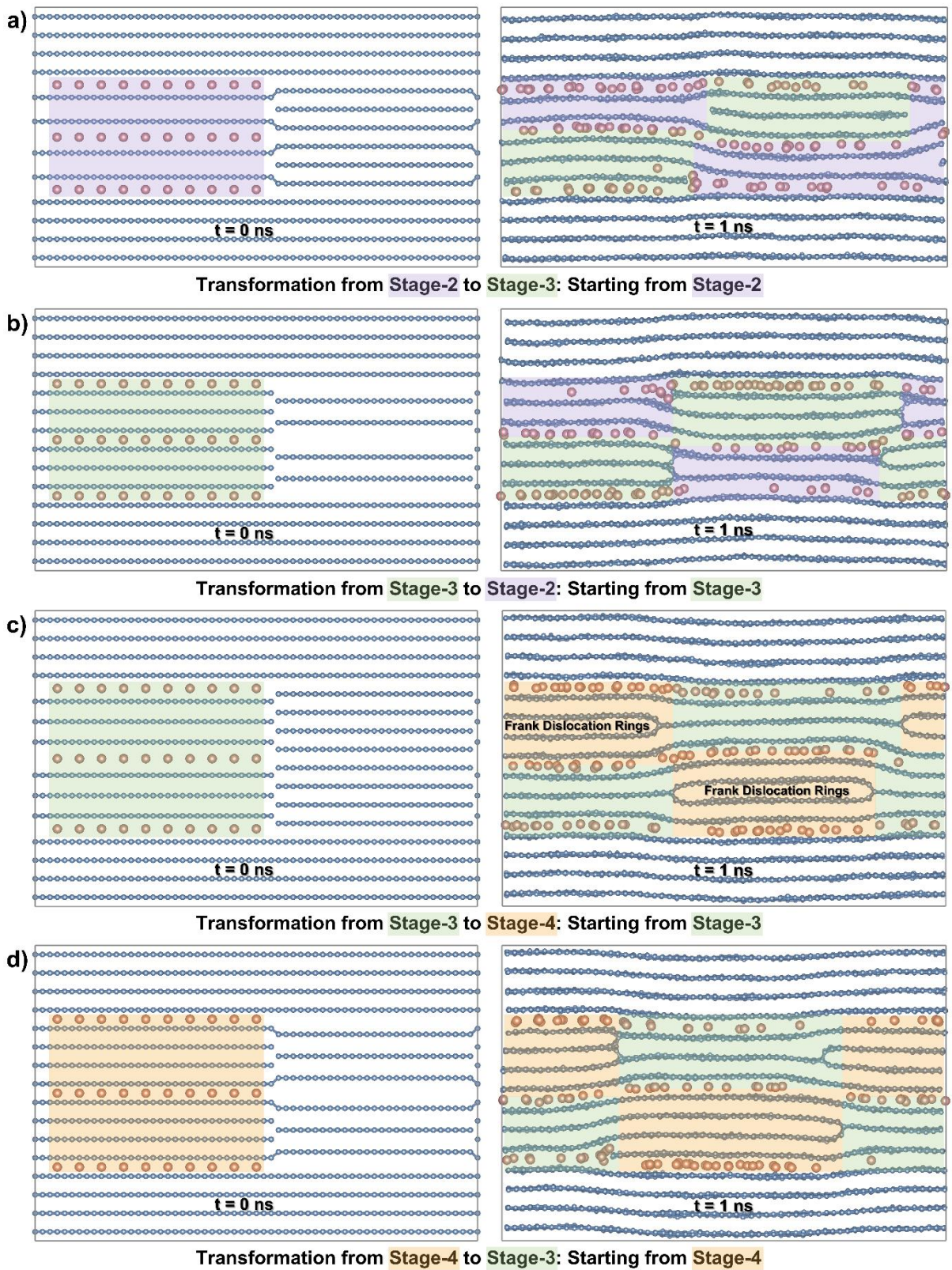

Figure 2: The impact of C layer sliding and reorganization on stage structure transitions during the Li insertion/extraction process. (a) The combined model of Stage-2 structure and graphite structure evolves into Stage-3 structure. (b) The combined model of Stage-3 structure and graphite structure evolves into Stage-2 structure. (c) The combined model of Stage-3 structure and graphite structure evolves into Stage-4 structure.

(d) The combined model of Stage-4 structure and graphite structure evolves into Stage-3 structure.

For the combined model of the Stage-2 structure and graphite structure with two additional C layers, as shown in Fig. 2a, after structural evolution for up to 100 ps, some regions retained the Stage-2 structure (purple background)[41]. Meanwhile, because the graphite side contains two extra C layers, local reorganization occurred during the evolution process, leading to the appearance of the Stage-3 structure (green background). Correspondingly, for the combined model of the Stage-3 structure and graphite structure with two fewer C layers, as shown in Fig. 2b, after 100 ps of structural evolution, some regions retained the Stage-3 structure (green background), while on the side where the Stage-3 structure has two additional C layers, local reorganization also occurred during the evolution, resulting in the emergence of the Stage-2 structure (purple background). The overall structural evolution shown in Figs. 2a and 2b demonstrates that the Stage-2 and Stage-3 structures can be interconverted through the local reorganization of C layers within the structure. At the same time, the presence of excess C layers can lead to the formation of staggered connections between C layers such as combinations of 3 and 2 layers.

Similarly, the Stage-3 and Stage-4 structures can also interconvert through the local reorganization of excess C layers within the structure, accompanied by the formation of staggered connections between C layers and Frank dislocation rings (text annotations in Fig. 2c). As shown in Fig. 2c, the combined model of the Stage-3 structure and graphite structure with two additional C layers generated Stage-4 regions after 100 ps of evolution. Meanwhile, as shown in Fig. 2d, the combined model of the Stage-4 structure and graphite structure with two fewer C layers formed Stage-3 regions after 100 ps of evolution.

Examining the interconversion processes between Stage-2 and Stage-3 shown in Figs. 2a-b, and between Stage-3 and Stage-4 shown in Figs. 2c-d, a common feature emerges: during the structural evolution, the local reorganizing direction of the C layers is always from regions with more C layers toward regions with fewer C layers, ultimately leading to an averaging of the C layer number in the final structure. This direction is driven by the goal of reducing internal stress within the structure. However, under our simulation conditions, we did not observe interconversion between Stage-1 and Stage-2 structures via C layer local reorganization. Combined with the structural evolution results shown in Figs. S2a-b, we speculate that the reason hindering the transition is that, in the combined models of Stage-1 or Stage-2 with graphite, the difference in C layer numbers on both sides is too great, resulting in excessive internal stress that makes C layer reorganizing difficult.

Moreover, based on the fact demonstrated in Fig. 1 and S7 that Stage-2 and Stage-3, as well as Stage-3 and Stage-4 structures, can transform into each other through C layer local reorganization, whereas Stage-1 and Stage-2 cannot, we speculate that the transitions between Stage-2, Stage-3, and Stage-4 reported in previous studies are solid-solution reactions (characterized by continuous changes in average C layer spacing), whereas the transition between Stage-1 and Stage-2 is a two-phase reaction (characterized by abrupt changes in average C layer spacing). The possible source of this difference lies in whether structural adjustments can be achieved internally through C layer local reorganization to facilitate stage structure transitions, rather than relying on external Li insertion/extraction processes.

From the perspective of LIBs applications, the mechanism revealed in our simulations provides a pathway for stress release and structural reconfiguration in graphite anodes during practical charge and discharge processes. This intrinsic structural buffering mechanism enables the graphite anode to adaptively adjust its internal layers under high-rate cycling, thereby reducing the risk of local stress concentration and structural degradation, and improving structural reversibility and stability. In contrast, the difficulty of achieving transitions between Stage-1 and Stage-2 via local reorganization means that, under rapid charge/discharge conditions, the system is more likely to form two-phase regions, resulting in pronounced stress concentrations and subsequently triggering failure modes such as conductive particle fracture. This explains the microscopic origins for the capacity fade and shortened lifespan of graphite anodes under high-rate operation. Accordingly, the strategies including regulating the stacking fault density to lower local reorganization energy barriers and utilizing operational protocols to avoid high-stress phase transitions are expected to enhance reversibility.

## 2. Asymmetric Charge/Discharge Process

After investigating the stage structure transitions of Li-GICs, we shifted our focus to performing full charge/discharge process simulations, aiming to explore the potential atomic-scale mechanisms underlying the asymmetric charge/discharge curves observed in experiments. We simulated the structural evolution during charge/discharge for both defect-free C layer structures and one-atom-missing C layer structures. By tracking the changes in the mean square displacement (MSD) of Li, the MSD of C, and the concentration of each Li layer, we analyzed the changes in Li-ion transport and structural evolution during the Li insertion/extraction process.

For the entire Li insertion process in the defect-free C layer structure, as shown in Fig. 3a, we use the state of charge (SOC) to describe the progress percentage of the process. The trends in Li MSD and C MSD indicate that as lithium continues to be inserted, the MSD of Li gradually decreases (from ~850 Å² to ~60 Å²), which is attributed to the increased hindrance to Li mobility as the Li concentration in the structure rises. Meanwhile, the MSD of C remains at approximately 60 Å² throughout the entire Li insertion process, indicating significant sliding of the C layers within the structure. Based on the structural snapshots in the figure, we speculate that Li ions between the C layers have an anchoring effect on the C layers, and the significant sliding is mainly due to the uneven distribution of Li during the insertion process, which causes some C layers to be less strongly anchored. Additionally, the changes in Li concentration across the layers, shown in Fig. S3a, suggest that Li tends to continuously insert into unsaturated Li layers, indicating a certain level iof selectivity in the insertion positions.

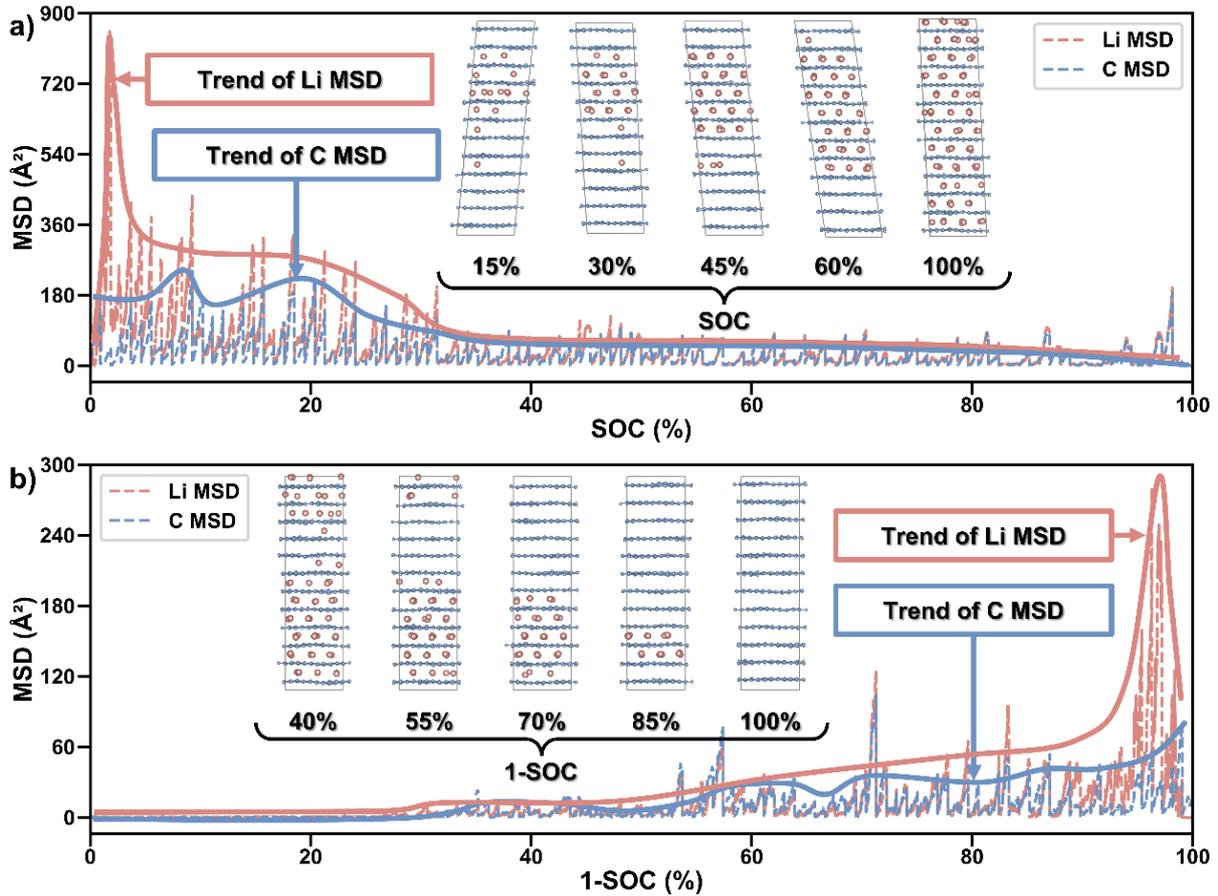

Figure 3: Simulation of the charge and discharge processes in the normal structure (structure with defect-free C layers). (a) Trends of Li MSD and C MSD with the progress percentage of Li insertion (represented by SOC), along with structural snapshots. (b) Trends of Li MSD and C MSD with the progress percentage of Li extraction (represented by 1-SOC), along with structural snapshots.

For the entire Li extraction process in the defect-free C layer structure, as shown in Fig. 3b, we use 1-SOC to describe the progress percentage of the process. The trends in Li MSD and C MSD indicate that as Li continues to be extracted, the Li MSD increases. Unlike the Li insertion process, the C MSD remains close to zero during the early process of Li extraction (progress percentage <40%) and only reaches a higher value (~60 Å²) in the later process, indicating that C layer sliding is not noticeable in the early process but becomes significant later. The structural snapshots in the figure help explain this: the lack of noticeable C layer sliding in the early stages is due to the pinning effect caused by the presence of Li between the C layers. However, in the later process, unpinned C layers appear in the structure, leading to noticeable sliding. Additionally, the changes in Li concentration across the layers shown in Fig. S3b suggest that Li tends to be extracted continuously from unsaturated Li layers, and the extraction positions are also selective.

For the entire Li insertion process in the one-atom-missing C layer structures, as shown in Fig. 4a, the progress percentage is described using SOC. The trends in Li MSD and C MSD indicate that as Li continues to insert, the Li MSD decreases. Unlike the normal structure (Fig. 3a), the C MSD remains close to zero after the Li insertion progress exceeds 40%, only reaching a higher value (~80 Å²) in the early process, indicating that C layer sliding essentially disappears in the middle and late process of Li insertion. The structural snapshots help explain this phenomenon: there are two main reasons for the disappearance of C layer sliding in the middle and late process. First, the defective C layers combine with interlayer Li, which enhances the pinning effect. Second, the Li distribution in the structure is relatively uniform, with Li present between each pair of adjacent C layers, so there are no C layers prone to sliding. Additionally, the changes in Li layer concentrations shown in Fig. S4a display a significant fluctuation with an approximately linear increasing trend, indicating that Li can achieve interlayer transport via defects, thereby promoting a more uniform Li distribution within the structure.

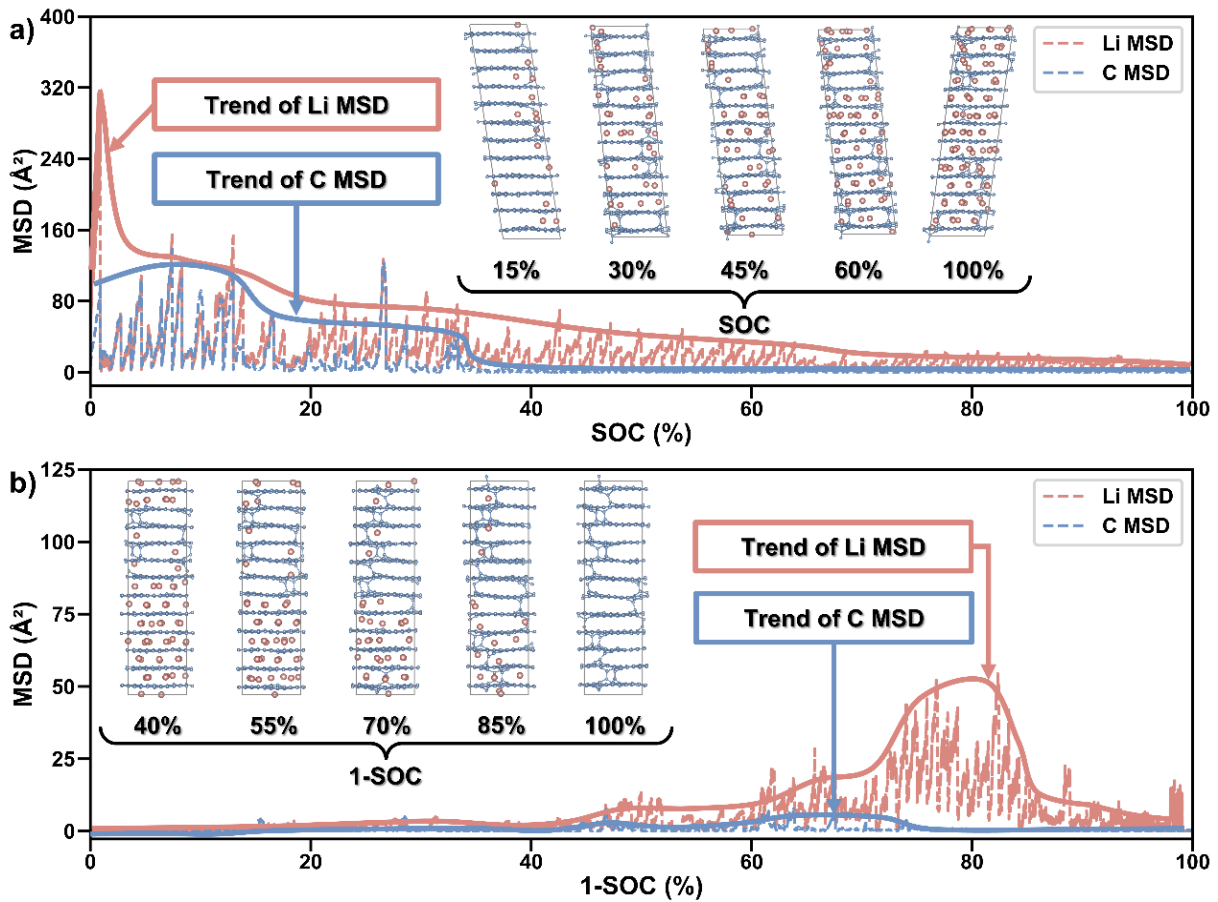

Figure 4: Simulation of the charge and discharge processes in the defect structure (structure with one-atom-missing C layers). (a) Trends of Li MSD and C MSD with the progress percentage of Li insertion (represented by SOC), along with structural snapshots. (b) Trends of Li MSD and C MSD with the progress percentage of Li extraction (represented by 1-SOC), along with structural snapshots.

For the entire Li extraction process in the one-atom-missing C layer structures, as shown in Fig. 4b, the progress percentage is described using 1-SOC. In terms of Li MSD and C MSD, unlike the normal structure (Fig. 3b), Li MSD first increases and then decreases as Li is continuously extracted, while C MSD remains at a low value close to zero throughout the entire extraction process. The structural snapshots help explain this phenomenon: the initial increase in Li MSD is due to the decrease in lithium concentration within the structure (which is consistent with the trend of Li MSD observed during the extraction process in the defect-free C layer structure), while the subsequent decrease is due to the interaction between Li and C layer defects, which restricts Li movement. When the 1-SOC exceeds 85%, most of the Li in the structure is associated with C layer defects, resulting in a reduction in Li MSD. The consistently low level of C MSD throughout the process is because all C layers are pinned by interlayer Li, so there are no C layers prone to sliding in the structure. Additionally, the changes in Li concentration across the layers shown in Fig. S4b exhibit an approximately linear decreasing trend with some fluctuations, indicating that interlayer Li transport also occurs during the Li extraction process.

By analyzing the changes in Li MSD and C MSD presented in Figs. 3 and 4, it can be observed that, during both Li insertion and extraction processes, the maximum values of Li MSD and C MSD in the one-atom-missing C layer structures are smaller than those in the defect-free C layer structure. The lower maximum value of Li MSD is due to the more uniform Li distribution in the defect structure, while in the normal structure Li is distributed non-uniformly; in addition, defects can associate with Li and hinder its transport. The lower maximum value of C MSD is because the uniform Li distribution and the association between defects and Li make the sliding of the C layers more difficult.

Our simulation results demonstrate pronounced asymmetry between lithium insertion and extraction processes, both in terms of Li transport capability and C layer sliding behavior, revealing distinct structural responses of layered graphite anodes during lithium intercalation and deintercalation. During the insertion steps, the non-uniform distribution of lithium leads to insufficient anchoring of certain C layers, making them more susceptible to sliding. This not only affects Li transport but may also increase polarization and cause fragmentation of conductive particles within the electrode material. In contrast, during the extraction steps, the residual lithium between layers initially exerts a pinning effect that effectively suppresses C layer sliding, with significant sliding only occurring when the Li content becomes extremely low. Fundamentally, this arises from the interactions between Li ions and the layered structure under different charge and discharge states, resulting in distinct structural dynamic responses. Meanwhile, from a battery application perspective, when an appropriate concentration of atomic-scale defects is present in the structure, the pinning effect produced by the association of lithium with defects can effectively suppress C layer sliding, thereby enhancing the reversibility and cycling stability of the electrode. However, an excessively high defect concentration may reduce Li-ion transport capability and thereby negatively impact battery performance.

### 3. Positional Selectivity of Li Insertion/Extraction and Its Implications

Based on the observation from full charge/discharge simulations that lithium tends to be consistently inserted into and extracted from unsaturated lithium layers, we further investigated the positional selectivity of lithium insertion/extraction.

For the positional selectivity in the Li intercalation process, we analyzed 4 transition processes: from graphite to Stage-8 (Fig. 5a), from graphite to Stage-4 (Fig. 5c), from Stage-4 to Stage-2 (Fig. 5e), and from Stage-2 to Stage-1 (Fig. 5g). To clarify the impact of Li concentration in unsaturated Li layers on positional selectivity, we considered unsaturated Li layer concentrations of 25%, 50%, and 75%, corresponding to low, medium, and high concentrations, respectively. For the Li intercalation positions, we primarily considered 3 types: unsaturated Li layers, unoccupied Li layers, and saturated Li layers. As shown in Figs. 5b and 5d, for the transitions from graphite to Stage-8 and from graphite to Stage-4, when the unsaturated Li layer is at low or medium concentration, most Li intercalation events occur in unsaturated Li layers, with fewer occurring in unoccupied Li layers. Only at high concentration does the probability of intercalation in unoccupied Li layers increase, but unsaturated Li layers still have the highest probability. Since saturated Li layers do not exist in the structure, the probability of intercalation events occurring in saturated Li layers is 0%. As shown in Figs. 5f and 5h, for the transitions from Stage-4 to Stage-2 and from Stage-2 to Stage-1, regardless of whether the unsaturated Li layer is at low, medium, or high concentration, most Li intercalation events still occur in unsaturated Li layers, with fewer occurring in unoccupied Li layers (the transition from Stage-2 to Stage-1 does not involve unoccupied Li layers), and none occur in saturated Li layers. Based on these results, the basic rules for Li intercalation positional selectivity can be summarized as follows: 1. Unsaturated Li layers are the primary positions for Li intercalation; 2. Saturated Li layers and unoccupied Li layers are unfavorable for Li intercalation, the former due to repulsive interactions with other Li ions, and the latter due to kinetic barriers caused by the lack of initial Li-C interactions; 3. Compared to unoccupied Li layers, Li intercalation in saturated Li layers is more difficult, and no intercalation events were observed in all simulations. According to these rules, we speculate that the Li intercalation process is largely regulated by the kinetics of C layers, where kinetically blocked unoccupied Li layers are opened by the intercalation of the first Li to become kinetically allowed unsaturated Li layers, and subsequently, Li continues to intercalate until saturation, while other unoccupied Li layers remain kinetically blocked, ultimately promoting the formation of various stage structures on a larger scale.

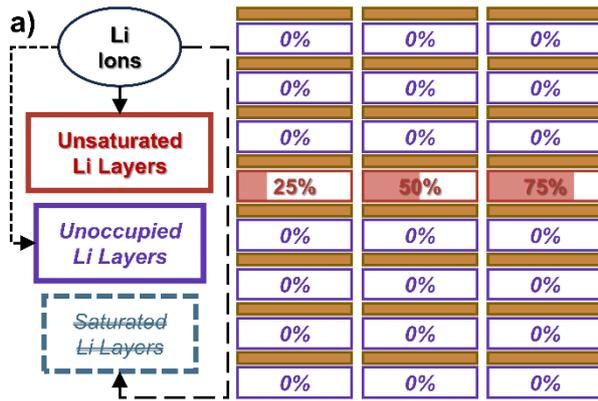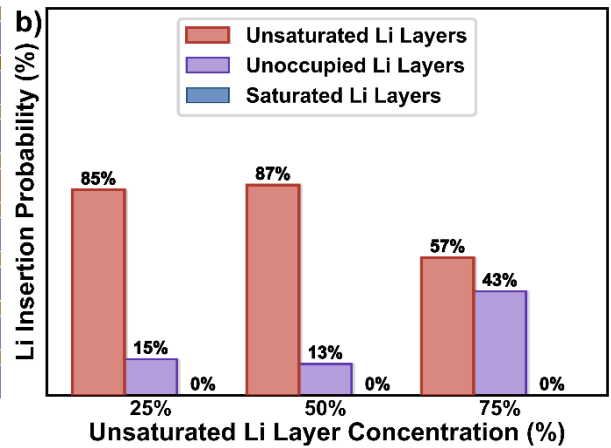

Li Insertion: from Graphite to Stage-8

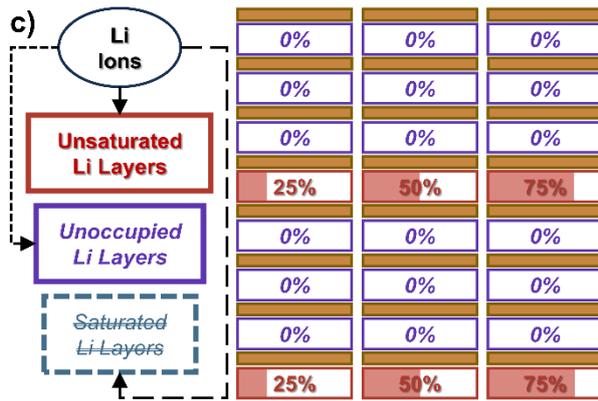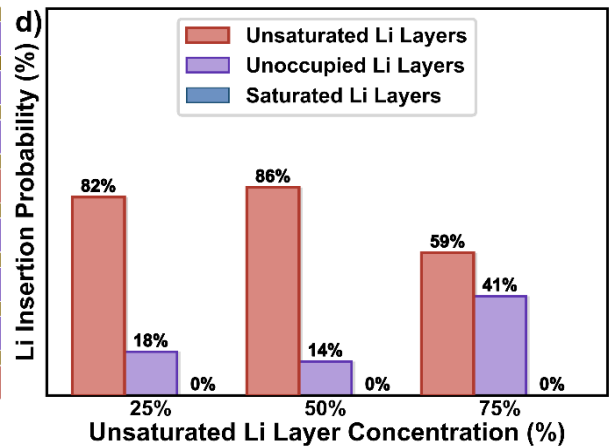

Li Insertion: from Graphite to Stage-4

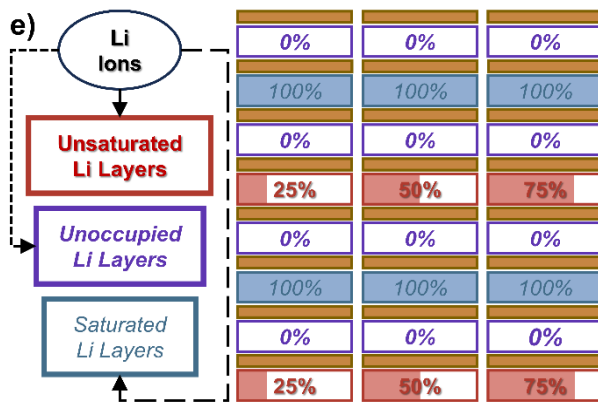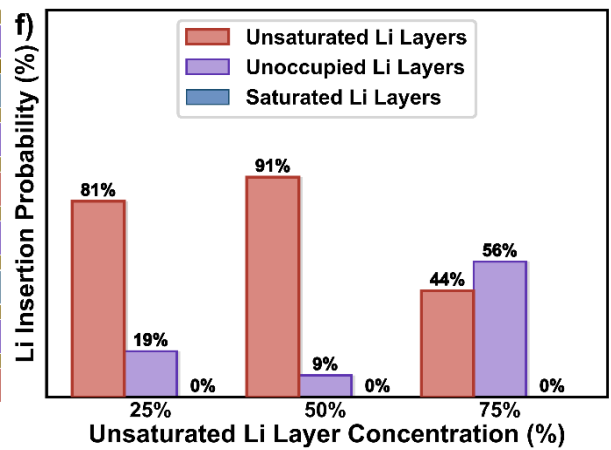

Li Insertion: from Stage-4 to Stage-2

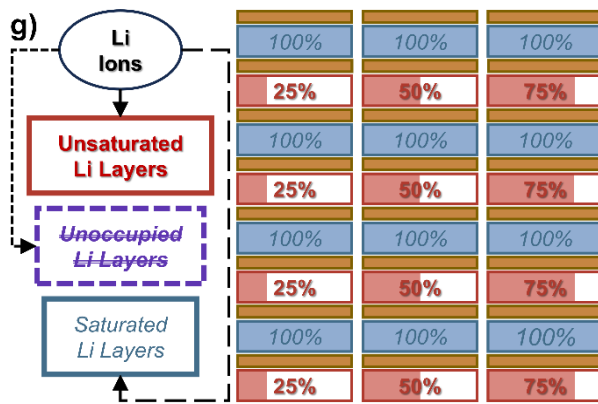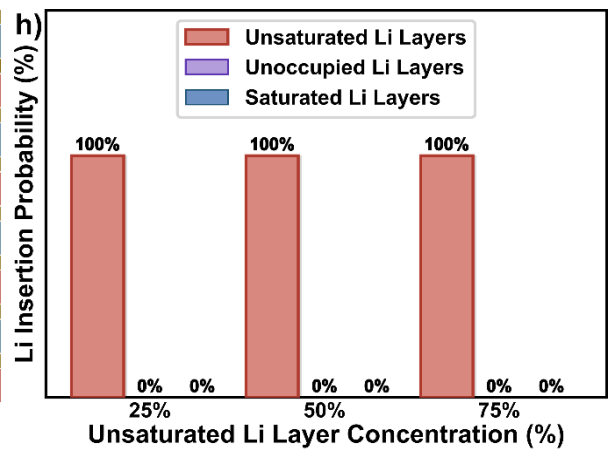

Li Insertion: from Stage-2 to Stage-1

Figure 5: Positional selectivity of Li insertion due to C layer dynamics. For the transitions (a) from graphite to stage-8, (c) from graphite to stage-4, (e) from stage-4 to stage-2, and (g) from stage-2 to stage-1, the Li distribution and available Li insertion positions in structural models at different unsaturated Li layer concentrations are shown (3 types: unsaturated Li layers, unoccupied Li layers, and saturated Li layers). For the transitions (b) from graphite to stage-8, (d) from graphite to stage-4, (f) from stage-4 to stage-2, and (h) from stage-2 to stage-1, the probability of Li insertion events occurring at the 3 types of available insertion positions at different unsaturated Li layer concentrations is statistically analyzed and presented.

For the positional selectivity in the Li extraction process, we analyzed 3 transition processes: from Stage-1 to Stage-2 (Fig. 6a), from Stage-2 to Stage-4 (Fig. 6c), and from Stage-4 to Stage-8 (Fig. 6e). In this analysis, the concentration of unsaturated Li layers was also controlled at 25%, 50%, and 75%, corresponding to low, medium, and high concentration conditions, respectively. As shown in Figs. 6b, 6d, and 6f, regardless of whether the unsaturated Li layers are at low, medium, or high concentration, the probability of Li extraction events occurring in the unsaturated Li layers is higher than that in the saturated Li layers. Based on the results above, the fundamental rules for the positional selectivity of Li extraction can be summarized as follows: 1. Unsaturated Li layers are the primary positions for Li extraction; 2. The tendency for lower stage structures to transform into higher stage structures is more pronounced during the transition from Stage-1 to Stage-2, whereas for other transitions, this tendency is relatively weaker; 3. Compared to the significant positional selectivity observed in the Li insertion process, the positional selectivity in the Li extraction process is relatively weaker. Based on these findings, we speculate that the main difference in the physical mechanism of Li extraction compared to Li insertion lies in the fact that, although C layer dynamics still play an important role in extraction, there is no transition from kinetically blocked to kinetically allowed C layers in the structure but the differences in extraction behavior arise between saturated and unsaturated Li layers, making the positional selectivity in the Li extraction process less pronounced than in the insertion process. Nevertheless, due to the presence of positional selectivity in Li extraction, the unsaturated Li layers in the structure will preferentially complete Li extraction and transform into unoccupied Li layers, thereby promoting the transition of the structure from lower stage to higher stage.

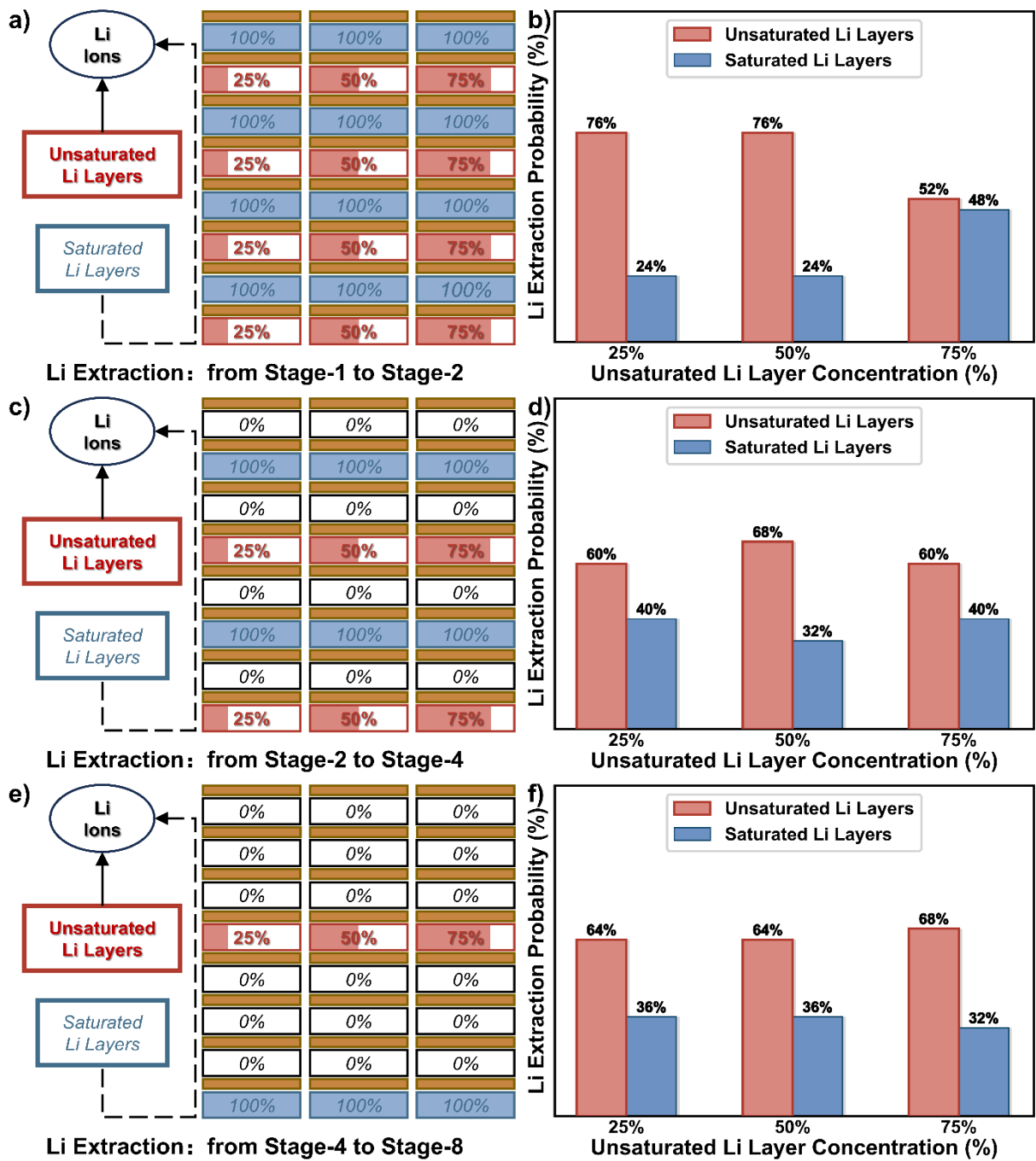

Figure 6: Positional selectivity of Li extraction due to C layer dynamics. For the transitions (a) from stage-1 to stage-2, (c) from stage-2 to stage-4, and (e) from stage-4 to stage-8, the Li distribution and available Li extraction positions in structural models at different unsaturated Li layer concentrations are shown (2 types: unsaturated Li layers and saturated Li layers). For the transitions (b) from stage-1 to stage-2, (d) from stage-2 to stage-4, and (f) from stage-4 to stage-8, the probability of Li extraction events occurring at the 2 types of available extraction positions at different unsaturated Li layer concentrations is presented.

Due to the positional selectivity of Li insertion and extraction, Li-GICs exhibit a series of structures with different lithium concentrations and distributions during the charge and discharge processes, accompanied by changes in Li transport capability and carbon layer dynamics. To clarify these effects, we further investigated the dynamics of Li layers using DPMD by statistically analyzing and comparing Li MSD and C MSD over 100 ps at 1000 K. As shown in Fig. 7a, we introduced varying numbers of low-concentration Li layers (with a lithium concentration of 25%) into structural models containing 1752-1914 atoms, while maintaining a 100% concentration in the remaining Li layers. The results show that the MSD of Li increases with the number of low-concentration Li layers.

For structures with the same Li concentration in each layer, we investigated the effect of uniformly changing Li concentration on MSD. As shown in Fig. 7b, as the Li concentration in the structure (~2000 atoms) decreased from 100% to 25%, the Li MSD increased by more than 7 times. The results from Figs. 7a and 7b indicate that a decrease in concentration reduces the restrictions on Li movement, thereby increasing the MSD, and vice versa.

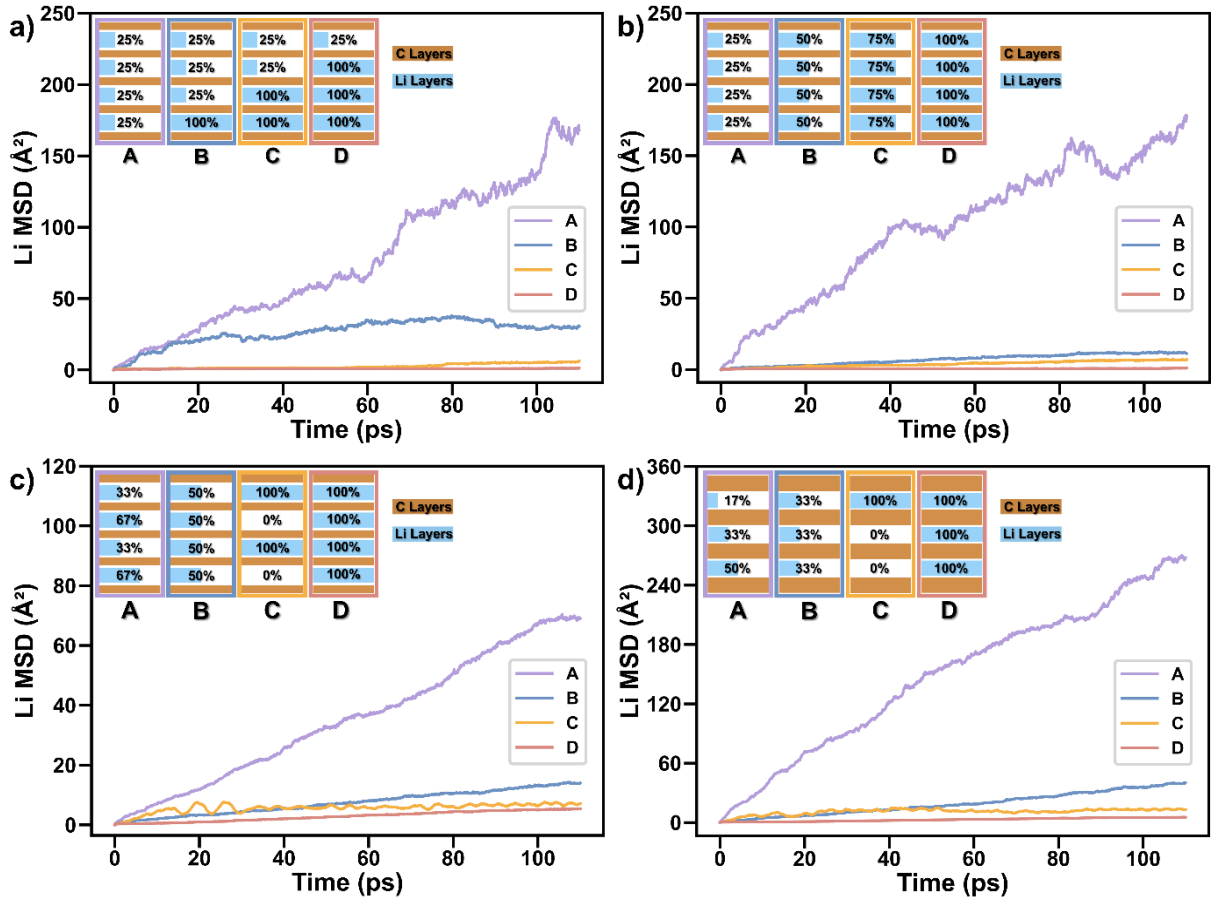

Figure 7: Effect of Li concentration and distribution on Li transport. (a) Trend of Li MSD with the number of low-concentration Li layers, where low-concentration Li layers have a concentration reduced to 25%, and the other Li layers maintain 100% concentration. (b) Relationship between Li MSD and Li concentration,

with all Li layers having the same concentration. Li MSD in (c) $LiC_{12}$ and (d) $LiC_{18}$ structures with non-uniform distribution, uniform distribution, and stage-structured distribution. Note: For panels c-d, the control group is the Stage-1 structure of $LiC_6$. The frame colors and letters in the insets of panels a-d correspond to the colors in the legend and the plot lines.

For structures with the same chemical composition but different Li distribution patterns, we compared the Li MSD in $LiC_{12}$ and $LiC_{18}$ models (~5000 atoms) under non-uniform, uniform, and stage structure distributions ($LiC_{12}$ as Stage-2, $LiC_{18}$ as Stage-3), using the Li MSD of $LiC_6$ Stage-1 as a reference. As shown in Figs. 7c-d, the results indicate that for stage structure distributions, the Li MSD of $LiC_{12}$ and $LiC_{18}$ is low, similar to that of $LiC_6$ Stage-1, due to the Li layers being in a saturated state with 100% concentration, which limits Li movement. However, both uniform and non-uniform distributions increase the Li MSD, with the effect being more significant in the non-uniform distribution. This is because the lower concentration Li layers contribute more to the Li MSD, and in non-uniform distribution, there are Li layers with even lower concentrations compared to the uniform distribution, leading to a greater Li MSD. These results on Li layer dynamics highlight the important role of Li concentration and distribution in regulating Li transport.

We also found that Li concentration and distribution influence the dynamics of the C layer (the ease of C layer sliding). As shown in Figs. S5a-d, as the Li concentration between C layers decreases, both the C MSD value and its fluctuation increase. In Figs. S3c-d, the C MSD in stage-structured distributions is notably higher than in other distributions. These results indicate that Li between the C layers indeed acts as a pinning force, and higher Li concentration strengthens this effect, making C layer sliding more difficult and reducing C MSD. Conversely, lower Li concentration weakens this pinning and increasing the C MSD.

The positional selectivity of Li insertion/extraction plays an important regulatory role in the transformation of various stage structures during the charge/discharge processes. Li preferentially inserts into or is extracted from unsaturated Li layers, which facilitates the sequential formation and evolution of different stage phases. Meanwhile, the differences in the underlying mechanisms and the degree of selectivity between Li insertion and extraction processes also contribute to the observed asymmetry between charge/discharge cycles. Furthermore, the positional selectivity of Li insertion/extraction leads to the formation of structures with a series of different Li concentrations and distributions, thereby affecting Li transport properties and the sliding behavior of C layers. These results suggest that optimizing the spatial distribution of Li, such as by creating locally low-concentration regions, can effectively enhance Li mobility and thus improve the rate capability of graphite anodes. At the same time, it is also important to maintain the structural stability of the C framework, such as by limiting C layer sliding through the pinning effect, which is essential for improving the mechanical stability and cycling life of graphite anodes.

4. **Defect-mediated regulation of Li transport and C layer dynamics**

Based on the effects of defects on Li transport, C layer sliding, and Li distribution observed in the full charge/discharge simulations, we recognize that defects also play a crucial regulatory role in Li-GICs. This regulatory effect is directly related to the type and spatial scale of the defects and thus needs to be discussed separately.

For atom-scale defects, we performed DPMD simulations at 1000 K for more than 50 ps on two structural models (~1200 atoms) with the same Li concentration and distribution: the normal structure with defect-free C layers and the defect structure with one-atom-missing C layers. As shown in Fig. 8a, compared with the random distribution of Li in the normal structure, Li in the defect structure tends to accumulate near the defects, indicating the formation of associations between Li ions and the defects. As shown in Fig. 5b, both the Li MSD and C MSD in the normal structure are higher than those in the defect structure, further confirming that the association between the defects and Li ions not only restricts Li mobility but also enhances the pinning effect on the C layers, thereby suppressing C layer sliding.

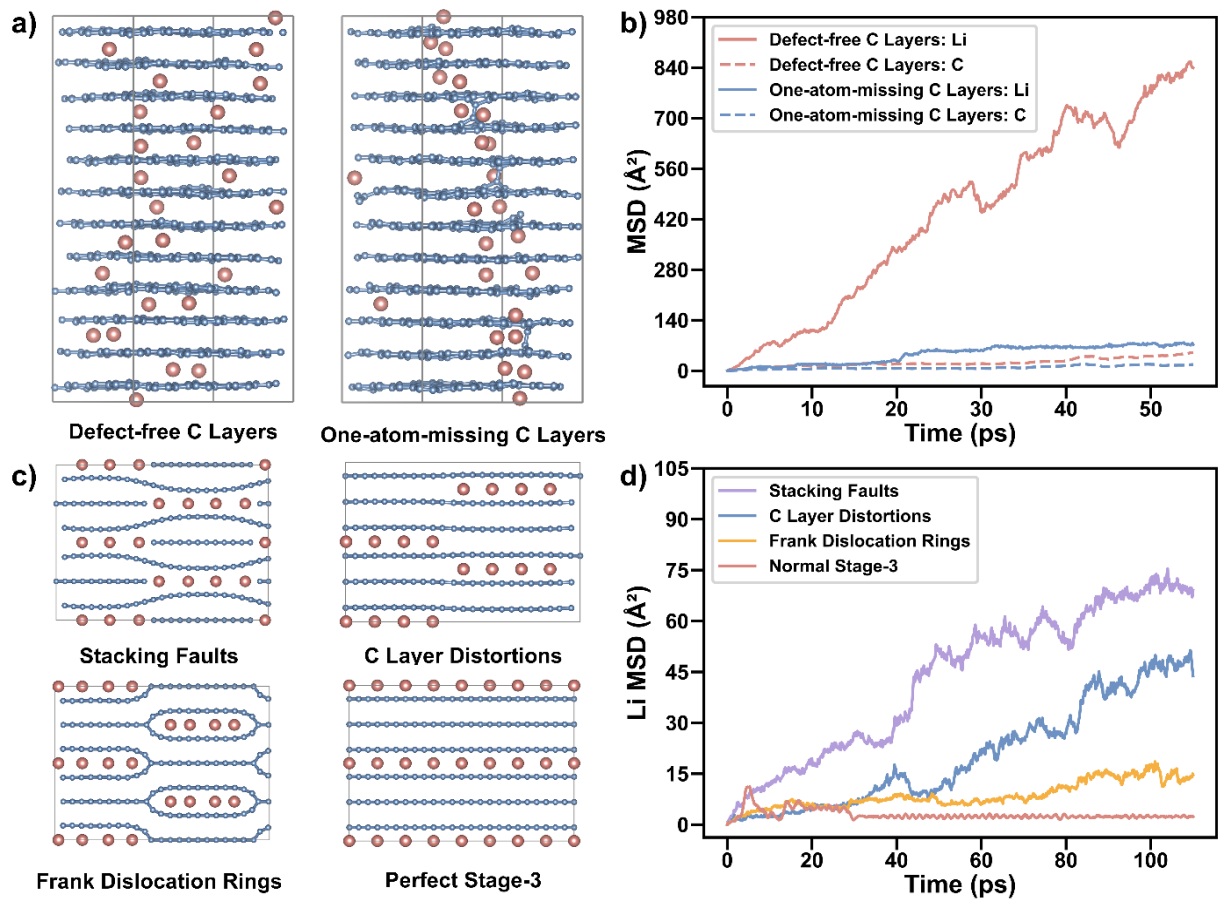

Figure 8: Impact of defects on Li transport and C layer dynamics in Li-GICs. (a) Tendency of Li accumulation near one-atom-missing defects. (b) Comparison of Li and C MSD between normal and defect structures. (c) Schematic representation of four models: normal Stage-3 structure and three types of entire C layer defects. (d) Comparison of Li MSD across structural models.

Apart from the atom-scale defects, Li-GICs may also develop defects that affect entire C layers. To investigate the impact of atom-layer-scale defects on Li transport and C layer sliding, we constructed four models (~1200 atoms), as shown in Fig. 8c: stacking faults, C layer distortions, Frank dislocation rings, and the normal Stage-3 structure. Similarly, we performed DPMD simulations at 1000 K for more than 100 ps on these four structural models. As shown in Fig. 8d, the Li MSDs of the three defect structures are all higher than that of the normal Stage-3 structure, indicating that atom-layer-scale defects can promote Li migration, and the extent of this promotion depends on the specific type of defect. For example, stacking faults and C layer bending have a more pronounced effect in enhancing Li transport, while the effect of Frank dislocation rings is weaker because the ring-shaped C layer structure constrains Li migration. Additionally, as shown in Fig. S6, the C layer MSD of the normal Stage-3 structure is significantly higher than those of the other three structures. This may be due to the presence of residual stress in the structures with atom-layer-scale defects, which further restricts C layer sliding compared to the normal structure.

From the perspective of practical battery applications, the presence and type of defects in graphite anodes are critically important for optimizing LIBs performance. Atom-scale defects can interact with Li ions, suppress Li migration and C layer sliding, which may hinder Li extraction and lead to irreversible capacity loss and decreased rate performance during cycling. In contrast, atom-layer-scale defects can typically enhance Li transport by providing additional migration directions and disrupting the ordered graphite structure, thereby improving rate capability and promoting the transformation of stage structures. However, if such defects become excessive, the additional stress introduced may compromise the structural stability of the electrode. Therefore, rationally controlling the type and concentration of defects in graphite anodes holds great promise for achieving an optimal balance between capacity retention, rate performance, and structural stability, thus enhancing the overall performance of practical LIBs.

**CONCLUSIONS**

In this work, we constructed a machine learning potential model for lithium-graphite intercalation compounds based on a deep potential model, and studied the dynamic evolution process of lithium-graphite intercalation compounds through deep potential molecular dynamics simulations. Firstly, we found that Stage-2, Stage-3, and Stage-4 structures can interconvert through the local reorganization of carbon layers whereas only the transition between Stage-1 and Stage-2 structures could not be realized, which suggests that the local reorganization of carbon layers provide a potential mechanism for stress release and structural self-adaptation in graphite anodes during practical charge-discharge processes, thereby enhancing the structural reversibility and stability. Secondly, through full-cycle charge-discharge simulations, we revealed the asymmetry between lithium transport capability and carbon layer sliding behavior during the lithium insertion/extraction processes, which originates from the distinct structural dynamic responses induced by the interactions between lithium ions and the layered structure under different charge and discharge states, while an appropriate concentration of defects can enhance the pinning effect to suppress carbon layer sliding and thereby improve the structural stability of the electrode. Subsequently, we investigated and verified the positional selectivity of lithium insertion/extraction processes, which tend to occur preferentially in unsaturated lithium layers, and found that such selectivity not only facilitates the formation and transformation of different staging structures but also leads to configurations with varying lithium concentrations and distributions, thereby revealing that the key to enhancing lithium transport capability lies in optimizing lithium distribution and maintaining low concentration layers, while simultaneously suppressing carbon layer sliding to maintain structural stability of the electrode. Finally, we investigated the regulatory effects of atomic-scale and atomic-layer-scale defects in carbon layers on lithium-ion transport and carbon layer sliding behavior and found that different types of defects exhibit varying impacts on lithium-ion transport capability while the presence of defects generally suppresses carbon layer sliding, thereby revealing that the critical role of controlling defect types and concentrations in simultaneously tuning lithium-ion transport performance and maintaining or enhancing the mechanical stability of graphite anode particles. This study can not only enable a deeper understanding of the dynamic processes and the continuous evolution of structure and properties in graphite anodes, but also offer a broadly applicable methodology that can inspire investigations of microstructural evolution in other electrode material systems.

**METHODS**

**DP Model Training, Testing and Further Validation**

1. Establishment of the Initial Dataset

The purpose of establishing the initial dataset is to provide the DP model with fundamental information on Li-GICs systems. Based on a comprehensive enumeration of possible scenarios that may occur during the structural evolution of Li-GICs, we constructed a series of models with various Li/C ratios, structural stages and defect types. Specifically, these models include: $C_{128}$ (128 atoms), $Li_1C_{128}$ (129 atoms), $Li_2C_{128}$ (130 atoms), $Li_4C_{72}$ (76 atoms), $Li_4C_{96}$ (100 atoms), $Li_4C_{104}$ (108 atoms), $Li_6C_{120}$ (126 atoms), $Li_8C_{96}$ (104 atoms), $Li_8C_{140}$ (148 atoms), $Li_8C_{144}$ (152 atoms), $Li_8C_{160}$ (168 atoms), $Li_8C_{192}$ (200 atoms), $Li_8C_{208}$ (216 atoms), $Li_9C_{180}$ (189 atoms), $Li_9C_{216}$ (225 atoms), $Li_{10}C_{110}$ (120 atoms), $Li_{12}C_{144}$ (156 atoms), $Li_{12}C_{240}$ (252 atoms), $Li_{16}C_{96}$ (112 atoms), $Li_{16}C_{288}$ (304 atoms, including stacking faults, atomic layer distortions, Frank dislocation loops, and normal Stage-3 structures), $Li_{18}C_{216}$ (234 atoms), $Li_{24}C_{144}$ (168 atoms), and $Li_{36}C_{216}$ (252 atoms). All systems were structurally optimized by density functional theory (DFT) calculations using the VASP package, employing the projector augmented-wave (PAW) method and the generalized gradient approximation (GGA) proposed by Perdew, Burke, and Ernzerhof (PBE). The cutoff energy (ENCUT) was set to 520 eV, the electronic convergence criterion was $1.0\times10^{-5}$ eV, and the ionic convergence criterion was -0.02 eV/Å. K-point sampling was based on the Gamma point with a grid density of $1\times1\times1$. To cover scenarios under various external pressures, the PSTRESS parameter during structural optimization was set to 0, 1, -1, 10, -10, 100, and -100. Meanwhile, to increase the randomness of structural evolution, the four types of $Li_{16}C_{288}$ structures were further subjected to over 20 ps of AIMD simulations under the NVT ensemble, and more than 1600 structures were selected based on their structural features to be included in the initial dataset. The temperature in the simulations was controlled using the Nose-Hoover thermostat, with simulated temperatures of 1000 K, 1250 K, and 1500 K. Other parameters such as k-point mesh, pseudopotential, functional, and cutoff energy were consistent with those used in structural optimization. In total, the resulting initial dataset contains DFT calculation results for approximately 15,000 structures.

2. Dataset Supplementation and Improvement

The purpose of supplementing the dataset is to improve the accuracy and generalizability of the DP model in describing the dynamic structural evolution of the Li-GICs system. Using the parallel learning framework Deep Potential Generator (DP-GEN), data generation was carried out iteratively, with each round consisting of three steps: training, exploration, and labeling[42].

*Exploration Step*

This step was carried out using LAMMPS combined with DPMD simulations[40]. The exploration temperatures included 50 K, 250 K, 500 K, 750 K, 1000 K, 1250 K, and 1500 K, while the pressures included -100 kbar, -10 kbar, -1 kbar, 0 kbar, 1 kbar, 10 kbar, and 100 kbar. The simulations were performed in the NPT ensemble, with a time step of 0.5 fs. One structure was sampled every 20 steps until the sampling process was completed.

Previous studies have shown that the maximum force model deviation set in DP-GEN can be used to screen all newly collected structures. Structures with model deviation values between the lower bound (σf_low) and the upper bound (σf_high) are subjected to DFT calculations in the subsequent labeling step, and the results are used as the exploration dataset. This dataset is then combined with the initial dataset to improve the accuracy and generalizability of the DP model in the next iteration. For the Li-GICs system, the initial settings were σf_low = 2.00 eV/Å and σf_high = 6.00 eV/Å. After the DP model performance improved, these values were reset to σf_low = 1.00 eV/Å and σf_high = 3.00 eV/Å for further optimization of the model.

*Labeling Step*

The main task in this step is to perform single ionic step calculations using VASP on the structures collected in the exploration dataset. The parameters for k-points, pseudopotential, functional, and cutoff energy were kept consistent with those used in structural optimization, and NSW was set to 1 to obtain the result for a single ionic step. Once the calculations for all structures in the exploration dataset were completed, the exploration dataset was merged with the initial dataset for use in the next round of DP-GEN iteration.

*Training Step*

In this step, the DP model was trained on the merged dataset comprising both the initial and exploration datasets. In each round, four models were trained using the DeePMD-kit package, with identical training parameters except for the random seeds. During training, mixed descriptors were used to simultaneously process information on interatomic distances and angles. Both the embedding network and the fitting network consisted of three hidden layers, with neuron numbers set to (25, 50, 100) and (240, 240, 240), respectively. The cutoff radius was set to 6 Å. The hyperparameters were: start pref_e = 0.02, limit pref_e = 1, start pref_f = 1000, limit pref_f = 1, start pref_v = 0.02, limit pref_v = 1. The loss function included not only energy and atomic forces but also the virial. The inclusion of the virial aims to ensure the accuracy of the description of internal interactions in the system, thereby guaranteeing the reliability of subsequent NPT ensemble DPMD simulations using the DP model. In each DP-GEN iteration, the DP model was trained for 1.5 million steps, with an initial learning rate of 0.001, decayed by 95% every 7,500 steps, and a final learning rate of $3.5 \times 10^{-8}$ at the end of training.

3. Improvement of Dataset Accuracy

The purpose of improving the dataset accuracy is to further enhance the reliability and precision of the DP model in describing the dynamic structural evolution of the Li-GICs system. Using VASP, we performed single ionic step DFT calculations with increased k-point density for all structures recorded in the initial dataset and in the exploration datasets generated from each iteration of dpgen. The parameters for pseudopotential, functional, and cutoff energy were kept consistent with previous calculations. The k-point density was still based on Gamma-point sampling. Through rigorous convergence tests, we found that when the number of k-points N in each direction satisfies the inequality cell_length × N > 28.5 (where cell_length is the lattice vector length in the a, b, or c direction, and N is the smallest positive integer satisfying the inequality), the total energy deviation of each structure converges within 1 meV/atom. Therefore, this criterion was applied to all structures to effectively balance computational accuracy, time, and cost.

4. Training and Testing of the Final DP Model

The dataset with improved accuracy obtained from step 3, containing DFT calculation results for approximately 30,000 structures, was used for the training of the final DP model. These 30,000 structures were divided into a training set and a test set in a 9:1 ratio, and the training set was used for model training with hyperparameters kept consistent with those in the training step described in section 2. The number of training steps was increased to 5 million, with an initial learning rate of 0.001 and a decay rate of 95% every 25,000 steps, resulting in a final learning rate of $3.5 \times 10^{-8}$ at the end of training. Upon completion of model training, the DeePMD-kit package was used to evaluate the model on the pre-divided test set. The final test results showed that the energy prediction MAE was 2.787 meV/atom and RMSE was 5.276 meV/atom (Fig. 1b); the atomic force prediction MAE was 125 meV/Å and RMSE was 238 meV/Å (Fig. 1c); and the virial prediction MAE was 6.188 meV/atom and RMSE was 10.39 meV/atom.

5. Further Validation of the DP Model

*Validation of the Accuracy of Average Interlayer Distance and Average C–C Bond Length Based on Structural Optimization*

We performed structural optimization calculations on graphite, AB-stacked Stage-4 structure ($Li_2C_{48}$), AB-stacked Stage-2 structure ($Li_2C_{24}$), AA-stacked Stage-4 structure ($LiC_{24}$), AA-stacked Stage-3 structure ($LiC_{18}$), AA-stacked Stage-2 structure ($LiC_{12}$), and AA-stacked Stage-1 structure ($LiC_6$) using both DFT and DP models. The average interlayer distance and average C–C bond length were then calculated from the optimized cell parameters. In all these structural models, the carbon layers are oriented perpendicular to the *c* lattice vector and parallel to the plane formed by the *a* and *b* lattice vectors (where the lengths of the *a* and *b* lattice vectors are equal and the angle between them is 120°). Thus, the average interlayer distance is defined as the length of the *c* lattice vector divided by the number of carbon layers in the unit cell, and the average C–C bond length is defined as the length of the *a* lattice vector divided by the number of C–C bonds along the *a* direction. The pseudopotentials, functionals, cutoff energies, and k-point density settings for DFT structural optimization were consistent with those in Section 3, except that the maximum number of ionic steps was set to 200 and the convergence criterion for ionic steps was -0.02 eV/Å. Structural optimization using the DP model was conducted with the Atomic Simulation Environment (ASE), allowing the cell parameters to relax freely. Upon comparison (Fig. 1d), the average interlayer distances and average C–C bond lengths predicted by the DP model showed good agreement in both values and trends with the DFT results, with a maximum error of only 0.05 Å.

*Validation of DPMD Simulation Accuracy for the Full Li Insertion/Extraction Process*

We conducted DPMD simulations using the DP model under the NPT ensemble at 1500 K (pressure = 0). For the charging and discharging processes, we sequentially extracted Li from a Stage-1 structure $Li_{32}C_{192}$ (empirical formula $LiC_6$) until pure $C_{192}$ was obtained, and sequentially inserted Li into $C_{216}$ until the Stage-1 structure $Li_{36}C_{216}$ (empirical formula $LiC_6$) was formed, respectively. During these two processes, a series of intermediate structures were sampled, and single-point DFT calculations (using the same parameters as in Section 3) were performed for each structure. By comparing the total energies predicted by the DP model and those calculated by DFT, we further validated the accuracy of the DP model.

For determining the positions for Li extraction, we adopted a bubble queue method. First, all lithium ions in the structure were identified, and then each lithium ion was sequentially removed, generating a series of new structures each with one less Li atom. The total energy of each new structure was calculated and compared, and the position corresponding to the lowest total energy was identified as the preferred Li extraction site.

For determining the positions for Li insertion, we adopted a grid search method. First, a dense three-dimensional grid was generated within the structure, and a Li ion was sequentially inserted at each grid point, resulting in a series of new structures, each with one additional Li atom. The DP potential was then used to calculate the total energy of each structure, and the position corresponding to the lowest total energy was selected as the preferred Li insertion site. To balance computational accuracy and cost, we set the maximum grid spacing in each direction to 0.4 Å.

The Li extraction process for $Li_{32}C_{192}$ was divided into 33 sub-steps (with Li stoichiometry decreasing from 32 to 0), while the Li insertion process for $C_{216}$ was divided into 37 sub-steps (with Li stoichiometry increasing from 0 to 36). For each sub-step, a DPMD simulation of 20,000 steps with a time step of 0.5 fs was performed. For every 1000 steps, a structure is sampled, and DFT calculations are performed to obtain the total energy, which is then compared with the predictions of the DP model.

The results demonstrated that (Fig. 1e), throughout the entire Li insertion/extraction process, the total energy values and their trends predicted by the DP model showed excellent agreement with the DFT calibration results. This confirms the accuracy of the DP model for simulating the charge and discharge processes of Li-GICs, even as the Li concentration continuously changes.

To clearly distinguish the sources of test data used for testing and further validating the performance of the DP model, we have specifically created a table, as shown in Tab. S1.

**Analysis of C Layer Local Reorganization to Facilitate Stage Structures Transition**

We performed a series of DPMD simulations with stacking fault structure models (6312-7368 atoms, ~75 Å×17 Å×44 Å) as shown in Fig. 2 and Fig. S2, at 1200K (pressure = 0, NPT ensemble) for 100000 steps with a timestep of 1fs. Structures were saved every 100 steps, and the information on structural evolution was recorded in a trajectory file to understand the structural evolution process and for comparative analysis.

**Full-cycle Charge/Discharge Simulations: for Normal Structure (Structure with Defect-free C Layers) and Defect Structure (Structure with One-atom-missing C Layers)**

The full-cycle charge/discharge simulation for the defect-free C layer structure in Fig. 3 is based on DPMD simulations of the entire Li insertion process for $C_{648}$ and the entire Li extraction process for $Li_{108}C_{648}$ (~13 Å×13 Å×40 Å). Because the stoichiometric number of Li varies from 0 to 108, the Li insertion/extraction process is divided into 109 sub-steps. Each sub-step involves running DPMD simulations for 35,000 steps at 1000K (pressure = 0, NPT ensemble) with a time step of 0.5 fs, and MSD information is output every 100 steps. After completing the entire Li insertion/extraction process, the MSD data from each sub-step is recalculated starting from the 5,000th step, concatenated in sequence, and the simulation time is converted into SOC (for Li insertion) or 1-SOC (for Li extraction) for plotting and analysis.

The full-cycle charge/discharge simulation for the one-atom-missing C layer structure in Fig. 4 is based on DPMD simulations of the entire Li insertion process for $C_{636}$ and the entire Li extraction process for $Li_{108}C_{636}$ (~13 Å×13 Å×40 Å). The division of sub-steps, simulation conditions, and data sampling and processing methods are the same as mentioned above.

For the statistical analysis of Li concentration in each Li layer within the structure, we first use the K-Means class from the sklearn module in Python to identify C layers in the structure through clustering analysis.[42] This provides the average coordinates of each C layer along the c direction (perpendicular to the C layers). Based on these average coordinates, we divide a series of intervals, and using these intervals and the distribution of Li coordinates in the c direction, we calculate the number of atoms in each Li layer. Dividing by the number of atoms when the Li layer is saturated gives the Li concentration in the layer. For Fig. S3 and S4, which describe the change in Li layer concentration with the percentage of simulation progress, the final structure obtained from each sub-step is used for statistical analysis.

For both structures, the locations for Li insertion and Li extraction were determined using the grid search method and bubble queue method, respectively (see *DP Model Training, Testing and Further Validation*, Section 5, Part 2: *Validation of DPMD Simulation Accuracy for the Full Li Insertion/Extraction Process*). The maximum grid spacing in each direction for the grid search method was set to 0.4 Å.

**Analysis of Li Insertion/Extraction Positional Selectivity**

We first conducted 120,000 steps of DPMD simulation at 1000K (pressure = 0, NPT ensemble) on a series of structural models (~200 atoms, ~9 Å×9 Å×28 Å) with Li distribution as shown in Figs. 5 and 6, with a time step of 1 fs. A structure was saved every 4000 steps, resulting in a trajectory file containing 30 structures. The first 5 structures represent the thermal equilibrium process, so we selected the remaining 25 structures for analysis.

For the Li insertion process, we categorized the insertion positions into 3 types: unsaturated Li layers, unoccupied Li layers, and saturated Li layers. The Li insertion probability for each type of position is calculated as: (Li insertion frequency for a specific type of Li layer / equivalent number of Li layers for that type of Li layer)×100%. For example, in the structure transition from Stage-4 to Stage-2, there are 2 equivalent unsaturated Li layers, 4 equivalent unoccupied Li layers, and 2 equivalent saturated Li layers. The Li insertion frequency is determined by performing one Li insertion for each of the 25 structures and tracking the insertion position, with the insertion position determined by the grid search method (see *DP Model Training, Testing and Further Validation*, Section 5, Part 2: *Validation of DPMD Simulation Accuracy for the Full Li Insertion/Extraction Process*). To enhance the accuracy of Li insertion location determination, the maximum grid spacing in each direction was reduced to 0.25 Å.

For the Li extraction process, we categorized the extraction positions into 2 types: unsaturated Li layers and saturated Li layers. The Li extraction probability for each type of position is calculated as: (Li extraction frequency for a specific type of Li layer / equivalent number of Li layers for that type of Li layer)×100%. The Li extraction frequency is determined by performing one Li extraction for each of the 25 structures and tracking the extraction position, with the extraction position determined by the bubble queue method (see *DP Model Training, Testing and Further Validation*, Section 5, Part 2: *Validation of DPMD Simulation Accuracy for the Full Li Insertion/Extraction Process*).

In the calculation of Li insertion/extraction frequencies, the statistical method for the number of Li atoms in each Li layer is the same as in the section *Full-cycle Charge/Discharge Simulations: for Normal Structure (Structure with Defect-free C Layers) and Defect Structure (Structure with One-atom-missing C Layers)*, with the statistical logic matched to the different simulation processes involved.

**Effects of Li Concentration and Distribution on Li Transport and C Layer Dynamics**

The MSD results in Figs. 7a-b and Figs. S5a-b are obtained from DPMD simulations conducted at 1000K (pressure = 0) using models consisting of approximately 2000 atoms (~25 Å×22 Å×27 Å) in the NPT ensemble. The distribution of Li in the structures and the Li concentration in each Li layer are shown in the insets of Figs. 7a-b. The total number of steps in the DPMD simulation is 120,000, with a time step of 1fs. From the 10,000th step onward (skipping the initial 10,000 steps for thermal equilibration), MSD information averaged by the number of atoms is calculated and output every 100 steps. After the simulation is completed, the MSD information from each collected structural model is plotted and compared.

The MSD results in Figs. 7c-d and Figs. S5c-d are obtained from models consisting of approximately 6000 atoms (~34 Å×33 Å×43 Å), with the same simulation conditions, MSD data sampling, and processing methods as mentioned above.

**Exploring the Effects of Defects on Li Transport and C Layer Dynamics**

Figs. 8a-b conducted DPMD simulations of structures with defect-free C layers and structures with one-atom-missing C layers (~1200 atoms, ~17 Å×15 Å×42 Å). Figs. 8c-d conducted DPMD simulations of stacking faults, C layer distortions, Frank dislocation rings and normal Stage-3 structures, all with the same Li/C ratio (~1200 atoms, ~17 Å×30 Å×20 Å). These simulations use the same simulation conditions, data collection, and processing methods as in the section *Li Layer Dynamics: Effects of Li Concentration and Distribution on Li Transport and C Layer Dynamics*.

**ACKNOWLEDGEMENTS**

This work was financially supported by the Strategic Priority Research Program of Chinese Academy of Sciences (grant no. XDB1040300), and the National Natural Science Foundation of China (grants no. 52172258). We acknowledge the National Supercomputer Center in Tianjin and Bohrium AI for Science Research Space Station for providing computational resources.

**REFERENCES**

1. Zheng, M.; You, Y.; Lu, J. Understanding Materials Failure Mechanisms for the Optimization of Lithium-Ion Battery Recycling. *Nat. Rev. Mater.* **2025**, *10 (5)*, 355-368.
2. Tan, S.; Borodin, O.; Wang, N.; Yen, D.; Weiland, C.; Hu, E. Synergistic Anion and Solvent-Derived Interphases Enable Lithium-Ion Batteries under Extreme Conditions. *J. Am. Chem. Soc.* **2024**, *146 (44)*, 30104-30116.
3. Yuan, S.; Cao, S.; Chen, X.; Wei, J.; Lv, Z.; Xia, H.; Chen, L.; Ng, R. B. F.; Tan, F. L.; Li, H.; Loh, X. J.; Li, S.; Feng, X.; Chen, X. Anion-Modulated Solvation Sheath and Electric Double Layer Enabling Lithium-Ion Storage From -60 to 80 °C. *J. Am. Chem. Soc.* **2025**, *147 (5)*, 4089-4099.
4. Märker, K.; Xu, C.; Grey, C. P. Operando NMR of NMC811/Graphite Lithium-Ion Batteries:

Structure, Dynamics, and Lithium Metal Deposition. *J. Am. Chem. Soc.* **2020**, *142 (41)*, 17447-17456.

5. Croguennec, L.; Palacin, M. R. Recent Achievements on Inorganic Electrode Materials for Lithium-Ion Batteries. *J. Am. Chem. Soc.* **2015**, *137 (9)*, 3140-3156.

6. Duffner, F.; Kronemeyer, N.; Tübke, J.; Leker, J.; Winter, M.; Schmuch, R. Post-Lithium-Ion Battery Cell Production and Its Compatibility with Lithium-Ion Cell Production Infrastructure. *Nat. Energy* **2021**, *6 (2)*, 123-134.

7. Konar, S.; Häusserman, U.; Svensson, G. Intercalation Compounds from LiH and Graphite: Relative Stability of Metastable Stages and Thermodynamic Stability of Dilute Stage Id. *Chem. Mater.* **2015**, *27 (7)*, 2566-2575.

8. Tao, L.; Xia, D.; Sittisomwong, P.; Zhang, H.; Lai, J.; Hwang, S.; Li, T.; Ma, B.; Hu, A.; Min, J.; Hou, D.; Shah, S. R.; Zhao, K.; Yang, G.; Zhou, H.; Li, L.; Bai, P.; Shi, F.; Lin, F. Solvent-Mediated, Reversible Ternary Graphite Intercalation Compounds for Extreme-Condition Li-Ion Batteries. *J. Am. Chem. Soc.* **2024**, *146 (24)*, 16764-16774.

9. Papaderakis, A. A.; Ejigu, A.; Yang, J.; Elgendy, A.; Radha, B.; Keerthi, A.; Juel, A.; Dryfe, R. A. W. Anion Intercalation into Graphite Drives Surface Wetting. *J. Am. Chem. Soc.* **2023**, *145 (14)*, 8007-8020.

10. Zhao, Y.; Zhang, Y.; Wang, Y.; Cao, D.; Sun, X.; Zhu, H. Versatile Zero-to Three-dimensional Carbon for Electrochemical Energy Storage. *Carbon Energy* **2021**, *3 (6)*, 895-915.

11. Zhao, W.; Zhao, C.; Wu, H.; Li, L.; Zhang, C. Progress, Challenge and Perspective of Graphite-Based Anode Materials for Lithium Batteries: A Review. *Journal of Energy Storage* **2024**, *81*, 110409.

12. Chae, S.; Choi, S.; Kim, N.; Sung, J.; Cho, J. Integration of Graphite and Silicon Anodes for the Commercialization of High-Energy Lithium-Ion Batteries. *Angew. Chem. Int. Ed.* **2019**, *59 (1)*, 110-135.

13. Wang, F.; Yi, J.; Wang, Y.; Wang, C.; Wang, J.; Xia, Y. Graphite Intercalation Compounds (GICs): A New Type of Promising Anode Material for Lithium-Ion Batteries. *Adv. Energy Mater.* **2013**, *4 (2)*, 1300600.

14. Yao, J.; Zhu, G.; Huang, J.; Meng, X.; Hao, M.; Zhu, S.; Wu, Z.; Kong, F.; Zhou, Y.; Li, Q.;

Diao, G. Si/Graphite@C Composite Fabricated by Electrostatic Self-Assembly and Following Thermal Treatment as an Anode Material for Lithium-Ion Battery. *Molecules* **2024**, *29 (17)*, 4108.

15. Oka, H.; Makimura, Y.; Uyama, T.; Nonaka, T.; Kondo, Y.; Okuda, C. Changes in the Stage Structure of Li-Intercalated Graphite Electrode at Elevated Temperatures. *Journal of Power Sources* **2021**, *482*, 228926.

16. Kim, H. S.; Hyun, J. C.; Choi, Y.; Ha, S.; Kang, D. H.; Heo, Y. H.; Kwak, J. H.; Yoon, J.; Lee, J. B.; Kim, J.-Y.; Jin, H.J.; Lee, J.; Lim, H.; Yun, Y. S. A New Perspective for Potassium Intercalation Chemistry in Graphitic Carbon Materials. *Energy Storage Materials* **2024**, *70*, 103514.

17. Kitamura, T.; Takai, S.; Yabutsuka, T.; Yao, T. Relaxation Stage Analysis of Lithium Inserted Graphite. *Journal of Physics and Chemistry of Solids* **2020**, *142*, 109440.

18. Dimiev, A. M.; Shukhina, K.; Behabtu, N.; Pasquali, M.; Tour, J. M. Stage Transitions in Graphite Intercalation Compounds: Role of the Graphite Structure. *J. Phys. Chem. C* **2019**, *123 (31)*, 19246-19253.

19. Insinna, T.; Bassey, E. N.; Märker, K.; Collauto, A.; Barra, A.L.; Grey, C. P. Graphite Anodes for Li-Ion Batteries: An Electron Paramagnetic Resonance Investigation. *Chem. Mater.* **2023**, *35 (14)*, 5497-5511.

20. Wu, X.; Song, B.; Chien, P.; Everett, S. M.; Zhao, K.; Liu, J.; Du, Z. Structural Evolution and Transition Dynamics in Lithium Ion Battery under Fast Charging: An Operando Neutron Diffraction Investigation. *Advanced Science* **2021**, *8 (21)*, 2102318.

21. Insinna, T.; Barra, A.L.; Grey, C. P. Overhauser Dynamic Nuclear Polarization of Lithiated Graphite Anodes: Probing Bulk and Surface Structures. *Chem. Mater.* **2025**, *37 (14)*, 5167-5182.

22. Li, Y.; Lu, Y.; Adelhelm, P.; Titirici, M. M.; Hu, Y.S. Intercalation Chemistry of Graphite: Alkali Metal Ions and Beyond. *Chem. Soc. Rev.* **2019**, 48 (17), 4655-4687.

23. Dresselhaus, M. S.; Dresselhaus, G. Intercalation Compounds of Graphite. *Advances in Physics* **2002**, *51 (1)*, 1-186.

24. Drüe, M.; Seyring, M.; Rettenmayr, M. Phase Formation and Microstructure in Lithium-Carbon Intercalation Compounds during Lithium Uptake and Release. *Journal of Power*

*Sources* **2017**, *353*, 58-66.

25. Weng, S.; Wu, S.; Liu, Z.; Yang, G.; Liu, X.; Zhang, X.; Zhang, C.; Liu, Q.; Huang, Y.; Li, Y.; Ateş, M. N.; Su, D.; Gu, L.; Li, H.; Chen, L.; Xiao, R.; Wang, Z.; Wang, X. Localized-domains Staging Structure and Evolution in Lithiated Graphite. *Carbon Energy* **2022**, *5 (1)*, e224.

26. Wan, W.; Wang, H. First-Principles Investigation of Adsorption and Diffusion of Ions on Pristine, Defective and B-Doped Graphene. *Materials* **2015**, *8 (9)*, 6163-6178.

27. Hofmann, U; Rüdorff, W. The formation of salts from graphite by strong acids. *Trans. Faraday Soc.* **1938**, *34*, 1017-1021.

28. Daumas, N; Herold, A. Relations between phase concept and reaction mechanics in graphite insertion compounds. *C. R. Acad. Sci.* **1969**, *268(5)*, 373-375.

29. Sole, C.; Drewett, N. E.; Hardwick, L. J. InSitu Raman Study of Lithium-Ion Intercalation into Microcrystalline Graphite. *Faraday Discuss.* **2014**, *172*, 223-237.

30. Zheng, T.; Dahn, J. R. Effect of Turbostratic Disorder on the Staging Phase Diagram of Lithium-Intercalated Graphitic Carbon Hosts. *Phys. Rev. B* **1996**, *53 (6)*, 3061-3071.

31. Babar, M.; Parks, H. L.; Houchins, G.; Viswanathan, V. An Accurate Machine Learning Calculator for the Lithium-Graphite System. *J. Phys. Energy* **2020**, *3 (1)*, 014005.

32. Yang, P.Y.; Chiang, Y.H.; Pao, C.W.; Chang, C.C. Hybrid Machine Learning-Enabled Potential Energy Model for Atomistic Simulation of Lithium Intercalation into Graphite from Plating to Overlithiation. *J. Chem. Theory Comput.* **2023**, *19 (14)*, 4533-4545.

33. Chen, Q.; Zhang, C.; Lin, L.; Xie, Q.; Xu, W.; Qiu, Y.; Lin, J.; Wang, L.; Peng, D.L. Electrochemically Induced High Ion and Electron Conductive Interlayer in Porous Multilayer Si Film Anode with Enhanced Lithium Storage Properties. *Journal of Power Sources* **2021**, *481*, 228833.

34. Graziano, G. Deep Learning Chemistry Ab Initio. *Nat. Rev. Chem.* **2020**, *4 (11)*, 564-564.

35. Wang, H.; Zhang, L.; Han, J.; E, W. DeePMD-Kit: A Deep Learning Package for Many-Body Potential Energy Representation and Molecular Dynamics. *Computer Physics Communications* **2018**, *228*, 178-184.

36. Zeng, J.; Zhang, D.; Lu, D.; Mo, P.; Li, Z.; Chen, Y.; Rynik, M.; Huang, L.; Li, Z.; Shi, S.; Wang, Y.; Ye, H.; Tuo, P.; Yang, J.; Ding, Y.; Li, Y.; Tisi, D.; Zeng, Q.; Bao, H.; Xia, Y.;

Huang, J.; Muraoka, K.; Wang, Y.; Chang, J.; Yuan, F.; Bore, S. L.; Cai, C.; Lin, Y.; Wang, B.; Xu, J.; Zhu, J.-X.; Luo, C.; Zhang, Y.; Goodall, R. E. A.; Liang, W.; Singh, A. K.; Yao, S.; Zhang, J.; Wentzcovitch, R.; Han, J.; Liu, J.; Jia, W.; York, D. M.; E, W.; Car, R.; Zhang, L.; Wang, H. DeePMD-Kit v2: A Software Package for Deep Potential Models. *The Journal of Chemical Physics* **2023**, *159 (5), 054801*.

37. Lu, D.; Wang, H.; Chen, M.; Lin, L.; Car, R.; E, W.; Jia, W.; Zhang, L. 86 PFLOPS Deep Potential Molecular Dynamics Simulation of 100 Million Atoms with Ab Initio Accuracy. *Compute. Phys. Commun.* **2021**, *259*, 107624.

38. Fu, F.; Wang, X.; Zhang, L.; Yang, Y.; Chen, J.; Xu, B.; Ouyang, C.; Xu, S.; Dai, F.; E, W. Unraveling the Atomic‐scale Mechanism of Phase Transformations and Structural Evolutions during (de)Lithiation in Si Anodes. *Adv. Func. Mater.* **2023**, *33 (37)*, 2303936.

39. Thompson, A. P.; Aktulga, H. M.; Berger, R.; Bolintineanu, D. S.; Brown, W. M.; Crozier, P. S.; in 't Veld, P. J.; Kohlmeyer, A.; Moore, S. G.; Nguyen, T. D.; Shan, R.; Stevens, M. J.; Tranchida, J.; Trott, C.; Plimpton, S. J. LAMMPS-a Flexible Simulation Tool for Particle-Based Materials Modeling at the Atomic, Meso, and Continuum Scales. *Compute. Phys. Commun.* **2022**, *271*, 108171.

40. Momma, K.; Izumi, F. VESTA: A Three-Dimensional Visualization System for Electronic and Structural Analysis. *J. Appl. Crystallogr.* **2008**, *41 (3)*, 653-658.

41. Zhang, Y.; Wang, H.; Chen, W.; Zeng, J.; Zhang, L.; Wang, H.; E, W. DP-GEN: A Concurrent Learning Platform for the Generation of Reliable Deep Learning Based Potential Energy Models. *Compute. Phys. Commun.* **2020**, *253*, 107206.

42. Pedregosa, F.; Varoquaux, G.; Gramfort, A.; Michel, V.; Thirion, B.; Grisel, O.; Blondel, M.; Louppe, G.; Prettenhofer, P.; Weiss, R.; Weiss, R. J.; Vanderplas, J.; Passos, A.; Cournapeau, D.; Brucher, M.; Perrot, M.; Duchesnay, E. Scikit-Learn: Machine Learning in Python. *Journal of machine learning research* **2011**, *12*, 2825-2830.

# Revealing the Staging Structural Evolution and Li (De)Intercalation Kinetics in Graphite Anodes via Machine Learning Potential

Liqi Wang[a, b], Xuhe Gong[a, c], Zicun Li[a], Ruijuan Xiao[a, b, *] and Hong Li[a, b, *]

a Institute of Physics, Chinese Academy of Sciences, Beijing, 100190, China.

b School of Physical Sciences, University of Chinese Academy of Sciences, Beijing 100049, China

c School of Materials Science and Engineering, Key Laboratory of Aerospace Materials and Performance (Ministry of Education), Beihang University, Beijing 100191, China.

*Corresponding author. Email: rjxiao@iphy.ac.cn, hli@iphy.ac.cn

Table S1: Composition of the data used for testing and further validation of the DP model

| Test Item | Corresponding Data Figure | Data Composition |
|---|---|---|
| Atomic force prediction accuracy | Fig. 1b | Test set, including various stacking stages and defect structures under different pressures and lithium concentrations (same below) |
| Structural energy prediction accuracy | Fig. 1c | Test set |
| Structural virial prediction accuracy | / | Test set |
| Characteristic lengths prediction accuracy | Fig. 1d | DFT structural optimization results of AB-stack graphite, AB-stack Stage-4, AB-stack Stage-2, AA-stack Stage-4, and multiple AA-stack Stage-4 structures |
| Equation of state test | Fig. S1 | DFT structural optimization results of AA-stack and AB-stack Stage-2 structures under different pressures |
| Dynamic lithium (de)intercalation | Fig. 1e | Single-ion-step DFT calculation results of configurations with different lithium concentrations collected from DPMD simulations |

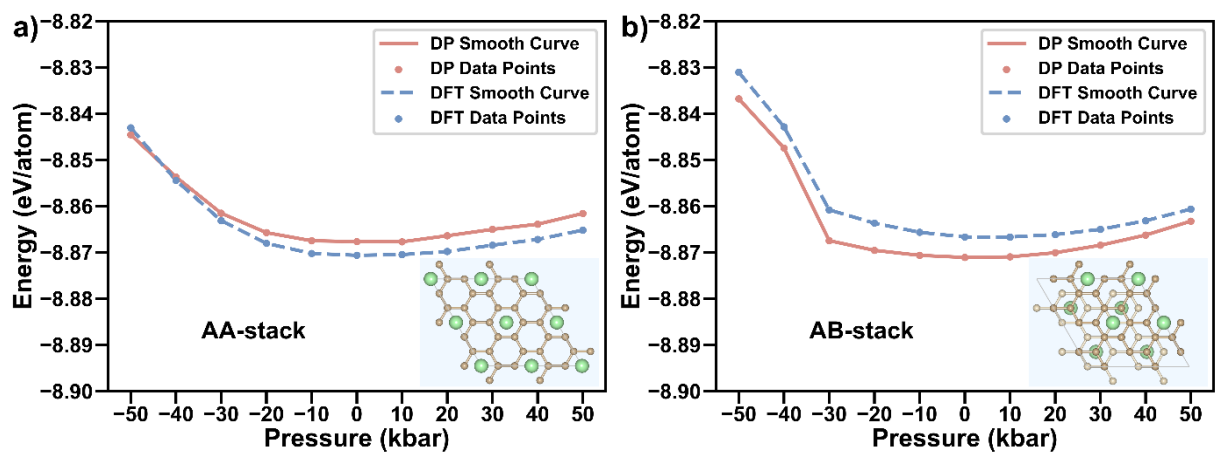

Figure S1: Thermodynamic validation of the DP model through equations of state (EOS) tests for (a)AA-stacked and (b)AB-stacked stage-2 structures under varying pressures.

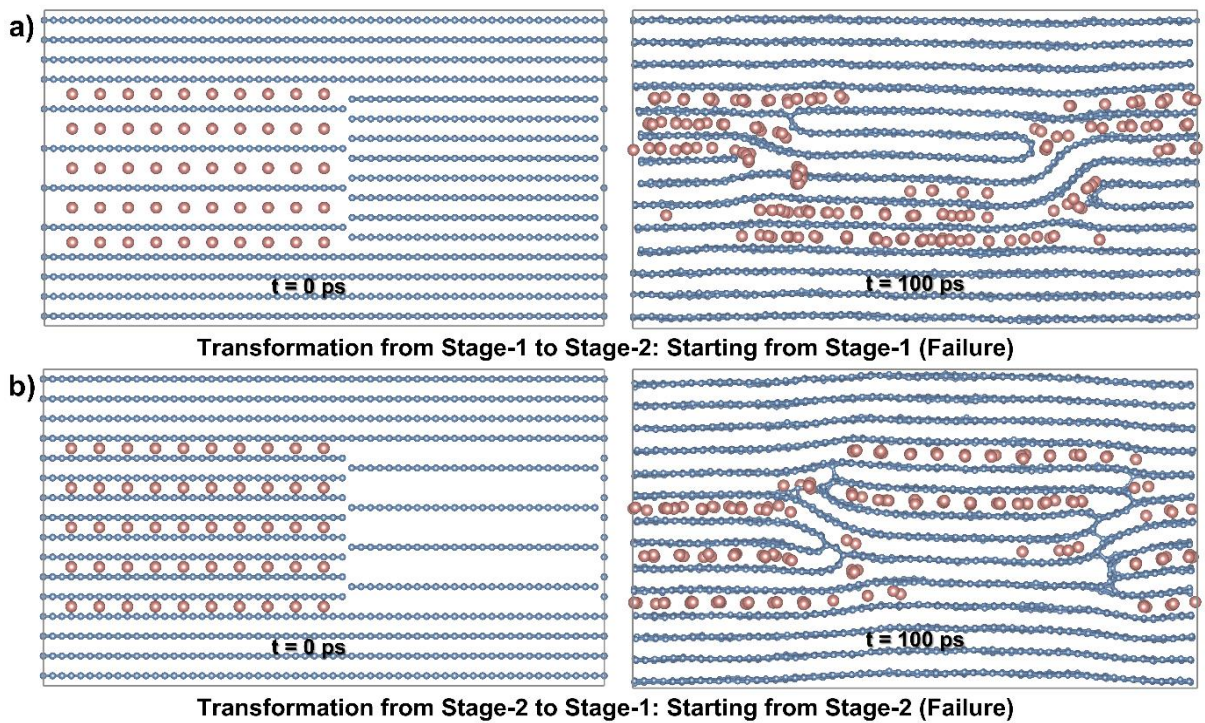

Figure S2: Attempts to achieve the mutual transformation between Stage-1 and Stage-2 structures starting from (a) Stage-1 and (b) Stage-2 configurations but failed.

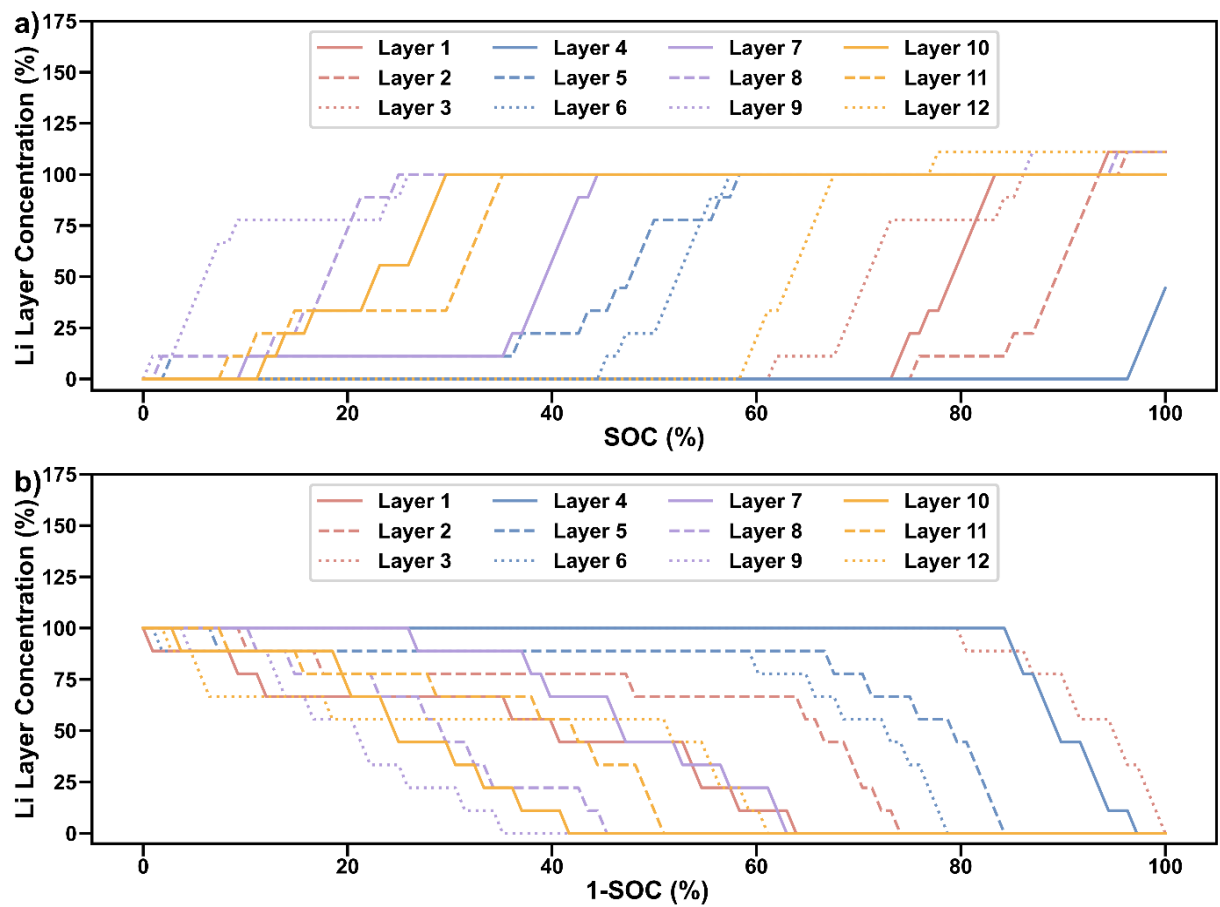

Figure S3: Variation of Li layer concentrations in the defect-free C layer structure throughout the process as a percentage of progress during (a) Li insertion and (b) Li extraction.

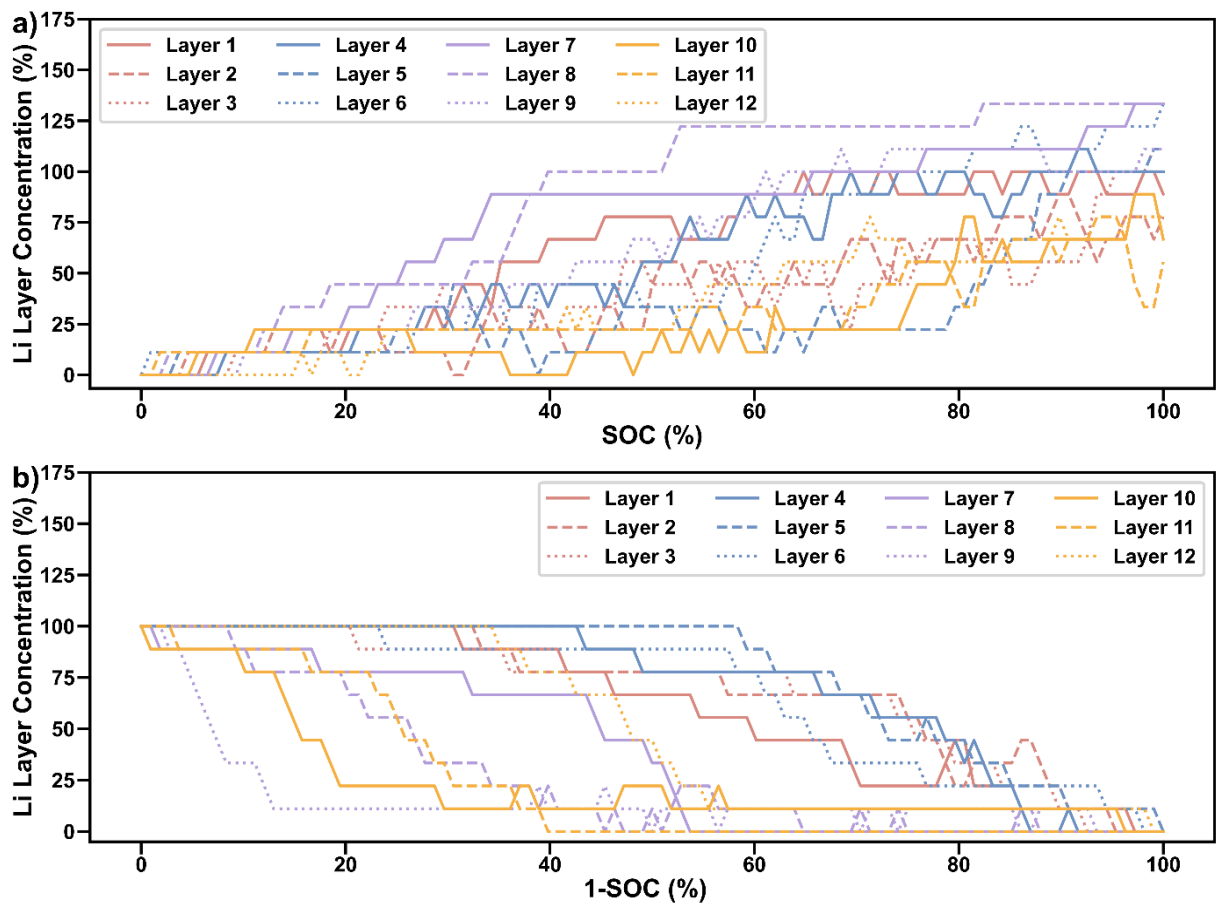

Figure S4: Variation of Li layer concentrations in the one-atom-missing C layer structures throughout the process as a percentage of progress during (a) Li insertion and (b) Li extraction.

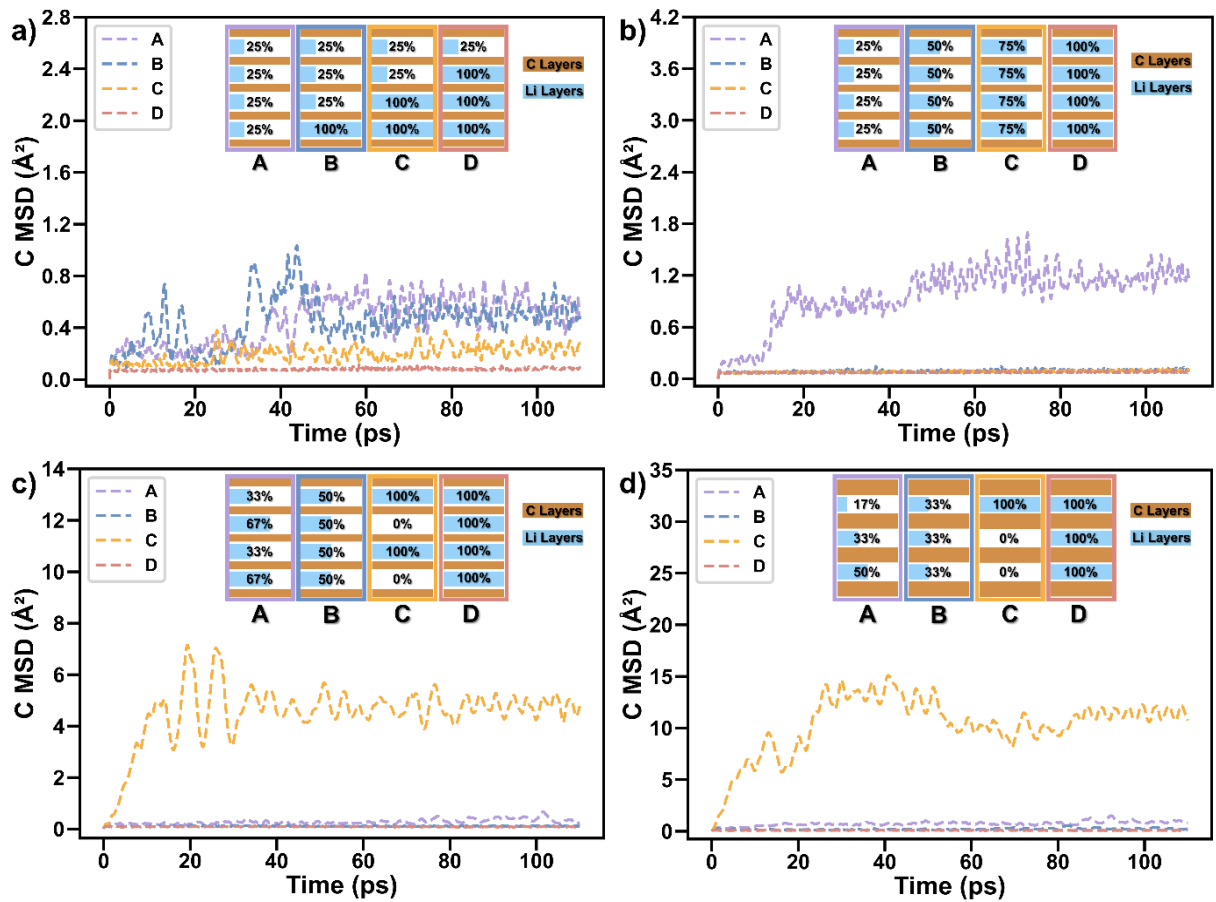

Figure S5: Effect of Li concentration and distribution on C layer dynamics. (a) Trend of C MSD with the number of low-concentration Li layers, where low-concentration Li layers have a concentration reduced to 25%, and the other Li layers maintain 100% concentration. (b) Relationship between C MSD and Li concentration, with all Li layers having the same concentration. C MSD in (c) $LiC_{12}$ structures with non-uniform distribution, uniform distribution, and stage-structured distribution, and in (d) $LiC_{18}$ structures with non-uniform distribution, uniform distribution, and stage-structured distribution. Note: For panels c-d, the control group is the Stage-1 structure of $LiC_6$. The frame colors and letters in the insets of panels a-d correspond to the colors in the legend and the plot lines.

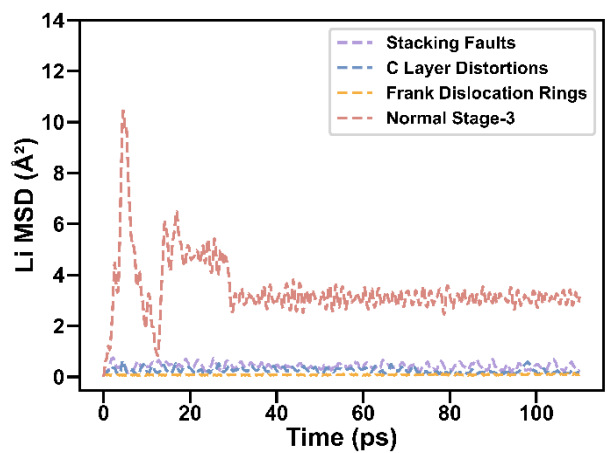

Figure S6: Variation of the C MSD over simulation time for 4 types of structures involving defects throughout the entire C layer.